\documentclass{aa}  
\usepackage{graphicx}
\usepackage{txfonts}
\usepackage{physics}
\usepackage[utf8]{inputenc}
\usepackage[english]{babel}
\usepackage[T1]{fontenc}
\usepackage{lmodern}
\usepackage{microtype}
\usepackage{booktabs}
\usepackage{bm}
\usepackage{graphicx}
\usepackage{amsfonts}
\usepackage{amsmath}
\usepackage{amssymb}
\usepackage{listings}
\usepackage[hidelinks]{hyperref}   
\usepackage{subfigure}
\usepackage{fancyhdr}
\usepackage{mathtools}
\usepackage{natbib}
\usepackage{comment}

\newcommand{\msun}{\mathrm{M}_\odot}
\newcommand{\plpeak}{\textsc{PowerLaw+Peak}}
\newcommand{\pobs}{p_\mathrm{obs}}
\newcommand{\pastro}{p_\mathrm{astro}}
\newcommand{\pdet}{p_\mathrm{det}}

\newcommand{\orcidicon}[1]{\href{https://orcid.org/#1}{\includegraphics[width=8pt]{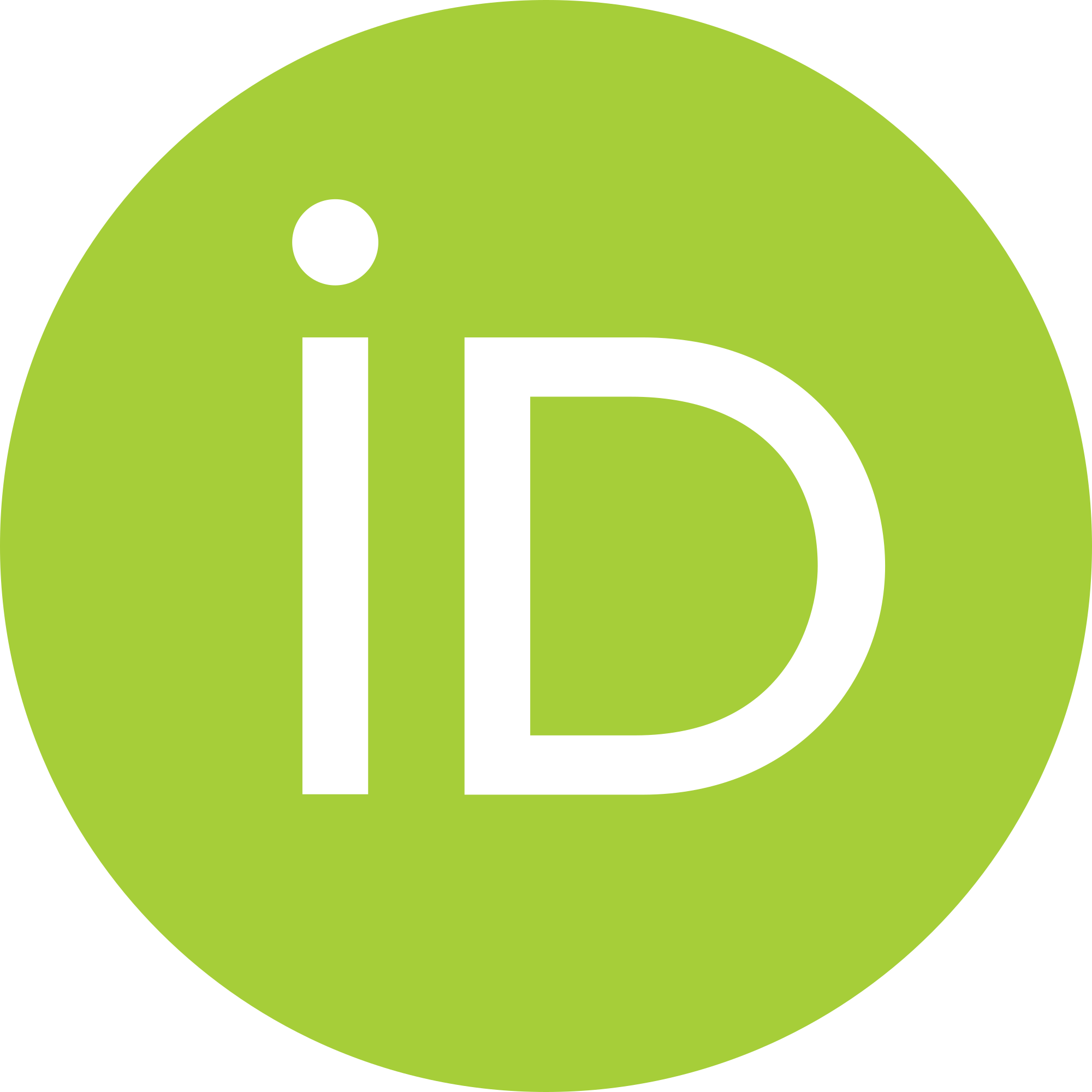}}}
\newcommand{\orcid}[1]{\href{https://orcid.org/#1}{\protect\orcidicon{#1}}}

\bibpunct{(}{)}{;}{a}{}{,} 

\begin{document} 

\titlerunning{Evolution of BH mass function with redshift}
\authorrunning{S.~Rinaldi et al.}
\title{Evidence for the evolution of black hole mass function with redshift}

\author{Stefano~{Rinaldi}\inst{1,2}\orcid{0000-0001-5799-4155}\thanks{E-mail: stefano.rinaldi@uni-heidelberg.de}\and  Walter~{Del~Pozzo}\inst{1,2}\orcid{0000-0003-3978-2030}\thanks{E-mail: walter.delpozzo@unipi.it}\and Michela~{Mapelli}\inst{3}\orcid{0000-0001-8799-2548}\thanks{E-mail: mapelli@uni-heidelberg.de}\and\\Ana~{Lorenzo-Medina}\inst{4}\orcid{0009-0006-0860-5700}\and Thomas~{Dent}\inst{4}\orcid{0000-0003-1354-7809}}

\institute{
    Dipartimento di Fisica ``E. Fermi'', Università di Pisa, Largo Bruno Pontecorvo 3, 56127 Pisa, Italy
    \and
    INFN, Sezione di Pisa, Largo Bruno Pontecorvo 3, 56127 Pisa, Italy
    \and
    Institut für Theoretische Astrophysik, ZAH, Universität Heidelberg, Albert-Ueberle-Str. 2, 69120 Heidelberg, Germany
    \and
    IGFAE, University of Santiago de Compostela, Rúa de Xoaquín Díaz de Rábago, 15782 Santiago de Compostela, Spain
    }

\date{Received \today; accepted XXX}
\abstract
{}
{We investigate the joint primary mass, mass ratio, and redshift observed distribution of astrophysical black holes using the gravitational wave events detected by the LIGO-Virgo-KAGRA collaboration and included in the third gravitational wave transient catalogue.}
{We reconstruct this distribution using Bayesian non-parametric methods, which are data-driven models able to infer arbitrary probability densities under minimal mathematical assumptions.}
{We find evidence for the evolution with redshift of both the primary mass and mass ratio distribution: our analysis shows the presence of two distinct sub-populations in the primary mass $-$ redshift plane, with the lighter population, $\lesssim 20\ \msun$, disappearing at higher redshifts, $z > 0.4$.
The mass ratio distribution shows no support for symmetric binaries.}
{The observed population of coalescing binary black holes evolves with look-back time, suggesting a trend in metallicity with redshift and/or the presence of multiple, redshift-dependent formation channels.}

\keywords{Stars: black holes -- Gravitational waves}

\maketitle

\section{Introduction}
GW150914 \citep{gw150914:2015}, the first gravitational wave (GW) event detected by the two LIGO detectors \citep{ligodetector:2015} during the first observing run (O1), was the first direct measurement of the intrinsic properties of a binary black hole (BBH). 
The two compact objects that composed this system were a surprise for the astrophysical community: before GW150914, stellar-sized black holes (BHs) were studied mainly via single-line spectroscopic binaries from the very first observation of the BH in the X-ray source Cygnus X-1 \citep{byram:1966,bahcall:1978}. The mass distribution for such objects was modelled as a Gaussian distribution centred at $7.8\ \msun$ with a standard deviation of $1.2\ \msun$ \citep{ozel:2010}. In contrast, the primary and secondary masses of GW150914, $36^{+5}_{-4}$ and $29^{+4}_{-4}\ \msun$ respectively \citep{GW150914_properties:2016,2021arXiv210801045T}, are significantly more massive than the maximum BH mass observed in Milky Way X-ray binaries \citep{farr:2011}. A common interpretation, although still controversial in its details, is that the  mass of a BH depends on the metallicity of its stellar progenitor, with metal-poor stars losing less mass by stellar winds and leading to the formation of more massive compact remnants than metal-rich stars \citep[e.g.,][]{heger:2003,mapelli:2009,mapelli:2010,zampieri:2009,belczynski:2010}.

Joined by the Virgo detector \citep{virgodetector:2015} during the second observing run (O2) in 2017, by KAGRA \citep{kagradetector:2013,kagradetector:2021} in 2020 at the end of the third run (O3), and expected to include LIGO-India \citep{ligo-india:2011} during the next decade, the LIGO-Virgo-KAGRA (LVK) collaboration detected, to date, almost 100 GW signals, most of them generated by BBH coalescences \citep{GWTC3:2021}. 
Since the LVK collaboration is now in the middle of the fourth observing run (O4), this number is expected to grow in the upcoming months.

These observations confirmed that the mass distribution inferred from Galactic BHs does not describe the general BBHs across the Universe, thus hinting that either coalescing BBHs follow different evolutionary paths from Galactic BHs, or they have a different origin altogether.  To discriminate among the various proposed scenarios, or unveil new ones, several groups have put considerable efforts towards the study of BBH populations \citep{astrodistGWTC1:2019,astrodistGWTC2:2021,astrodistGWTC3:2023,talbot:2018,stevenson:2019,edelman:2020,fishbach:2021,roulet:2021,galaudage:2021,bouffanais:2021,stevenson:2022,callister:2023,sadiq:2023}.

Current theoretical models propose several different formation channels for astrophysical BBHs. Among those, the most considered in the literature are isolated evolution 
\citep[e.g.,][]{bethe:1998,belczynski:2002,dominik:2012,eldridge:2016,stevenson:2017,klencki:2018,kruckow:2018,iorio:2023,fragos:2023,briel:2023,srinivasan:2023} and dynamical assembly  in star clusters \citep[e.g.,][]{portegieszwart:2000,banerjee:2010,rodriguez:2016,mapelli:2016,askar:2017,samsing:2018,banerjee:2018,dicarlo:2020,kumamoto:2020,kremer:2020,fragione:2020,arcasedda:2021,dallamico:2021,romero:2022} or  active galactic nucleus (AGN) discs \citep[e.g.,][]{mckernan:2012,mckernan:2018,bartos:2017,tagawa:2020,tagawa:2021,li2021,samsing:2022,vaccaro:2023}. Given the large observed masses in BBHs, the possibility that at least one of the two BHs composing the detected binary has already undergone a merger has also been put forward \citep{miller:2002,gerosa:2017,fishbach:2017:hier,rodriguez:2019,antonini:2019,kimball:2021,mapelli:2021,gerosa:2021,fishbach:2022,mould:2022,fragione:2023,rodriguez:2023}. 

The processes undergone by stars during their life are expected to leave signatures in the BH astrophysical distributions. For example, stellar winds \citep{vink:2001,lamers:2017}, core-collapse supernovae \citep{fryer:2012,schneider:2021}, envelope removal after common envelope \citep{hurley:2002}, and pulsational pair-instability supernovae \citep{heger:2003,woosley:2015} result in non-negligible mass loss for the star. In addition to this, there is also the possibility that the BH is not of astrophysical origin, like primordial BHs \citep{carr:1975}. 

The LVK collaboration explored the properties of BBH populations in a series of papers \citep{astrodistGWTC1:2019,astrodistGWTC2:2021,astrodistGWTC3:2023} employing different parametric and semi-parametric phenomenological models to investigate the BH properties. In particular, the BH mass distribution is inspired by the initial stellar mass function \citep{salpeter:1955,kroupa:2001} and revolves around a power-law model. Several additional features have been included throughout the years, such as a high mass hard cutoff and a power-law distribution for the mass ratio \citep{fishbach:2017,wysocki:2019}, or a Gaussian feature at around $35-40\ \msun$ \citep{talbot:2018}, resulting in the so-called `\plpeak' model. 
This model, described in App.~B.1b of \citet{astrodistGWTC3:2023}, is the fiducial population model for LVK analysis.

Models as such follow a phenomenological approach: they are inspired by the underlying physics, but capturing the many facets of astrophysical processes is  beyond their scope. Moreover, these models are constrained to a given functional form that may or may not represent the underlying population. As a consequence the inference based on them could be biased if some of the features of the BH distribution are not properly accounted for in the model.

Nonetheless, building a reliable parametric model that takes into account the complexity of astrophysical models is a formidable task: to date, and to the best of our knowledge, no all-encompassing parametric model has been proposed, although mixture models start to be explored. For instance, \citet{zevin:2021} show that single channel models are disfavoured compared to models that include multiple channels.

An alternative route for the inference of BBH astrophysics, based on an opposite approach to the all-encompassing modelling, is to use the so-called `non-parametric methods'. These methods allow to reconstruct probability densities from observations without being committal to any model-specific prescription. Being notably flexible and powerful, they are able to reconstruct arbitrary probability densities under minimal assumptions. Despite the potentially misleading name, these models do have parameters, either countably infinitely many or a flexible number of them. Their flexibility comes with the cost that no physical information can directly be inferred from the reconstructed distribution: any interpretation in terms of astrophysical processes and formation channels has to be done a posteriori.

Some works do explore this direction \citep{mandel:2016,tiwari:2018,edelman:2020,tiwari:2021:vamana,li:2021,toubiana:2023,ray:2023,sadiq:2023,callister:2023,edelman:2023} as well as some of the models included in \citet{astrodistGWTC3:2023}, however most of them are semi-parametric rather than non-parametric. The distinction among the two classes lies in the number of free parameters included in the model. Both classes include effective parameters that have the sole purpose of modelling the underlying distribution, meaning that a direct physical interpretation of them is not to be sought: for non-parametric models the number of such parameters is potentially infinite, allowing for a very large flexibility in modelling arbitrary distributions. Conversely, semi-parametric models include a finite number of effective parameters, and often the value of some of them has to be specified a priori. Thus, while being more flexible than parametric models, semi-parametric models often require non-negligible tuning of some of their parameters, for instance the maximum number of components in flexible mixture models or the typical width of the expected features.

In a previous publication, we proposed (H)DPGMM (short for `a hierarchy of Dirichlet process Gaussian mixture model'): a fully non-parametric multivariate method developed to investigate the BH mass distribution \citep{rinaldi:2022:hdpgmm}. In this study, we limited our investigations to the reconstruction of the one-dimensional distribution of the masses of BHs in merging binaries, essentially confirming what has been found by the LVK \citep{astrodistGWTC2:2021,astrodistGWTC3:2023}. In this paper, we extend our investigation to a multi-dimensional slice of the full BBH parameter space. We apply (H)DPGMM to the third GW transient catalogue (GWTC-3) to investigate the correlations among the primary mass of the binary $M_1$, the mass ratio $q = M_2/M_1$, and the redshift $z$. The presence of such correlations would be completely hidden within any one-dimensional -- hence marginal -- distribution. 
Spotting any correlations is paramount, since they could be the signature of different formation channels, shedding light on the processes that lead to the observed population of astrophysical BHs:
\citet{callister:2021} and more recently \citet{adamcewicz:2023} report the presence of an anti-correlation among $q$ and the effective spin $\chi_{\mathrm{eff}}$, whereas \citet{biscoveanu:2022} suggest that the spin distribution broadens with redshift. 

The paper is organised as follows: in Sec.~\ref{sec:methods}, we sketch the non-parametric method used in this work and discuss an improved method used here to account for selection effects over mass and redshift. 
In Sec.~\ref{sec:primarymass} we follow up the work of \citet{rinaldi:2022:hdpgmm}, reconstructing the primary mass distribution with GWTC-3; we then expand the parameter space in Sec.~\ref{sec:m1qz} to include the primary mass, mass ratio, and redshift, the main result of this paper.
Finally, we propose an astrophysical interpretation of such distribution in Sec.~\ref{sec:interpretation}. 

\section{Methods}\label{sec:methods}
The methods used in this work are largely based on a previous paper by the same authors \citep{rinaldi:2022:hdpgmm}, where we investigate the BBH mass distribution with a fully non-parametric approach using the data from the second GW transient catalogue (GWTC-2) \citep{GWTC2:2021}.
\citet{rinaldi:2022:hdpgmm} find that the non-parametric reconstruction of the BBH mass distribution is consistent with all the four parametric models used in \citet{astrodistGWTC2:2021}. 
Hereafter, we briefly outline the main features of (H)DPGMM, the hierarchical non-parametric method introduced by \citet{rinaldi:2022:hdpgmm}, based on the Dirichlet process Gaussian mixture model (DPGMM). The notation follows the one of the aforementioned paper. Details about the implementation and sampling scheme can be found in \citet{rinaldi:2022:figaro}. Details of both the model used and the inference scheme can be found in these papers. The notation of this section closely follows the one introduced in \citet{rinaldi:2022:hdpgmm}.

The basic idea of the DPGMM is that any arbitrary probability density $p(x)$ can be approximated as infinite weighted sums of Gaussian distributions \citep{nguyen:2020}:
\begin{equation}
    p(x)\simeq \sum_i^\infty w_i\,{} \mathcal{N}(x|\mu_i,\sigma_i)\,.
\end{equation}
This result holds also for multivariate probability densities, simply replacing the Gaussian distribution with its multivariate extension as kernel function: this paper is based on this feature\footnote{While dealing with parameters with boundaries, such as $q$, we make use of the parametrization presented in \citet{rinaldi:2022:hdpgmm}, Section 3.1., for each parameter separately.}.

The DPGMM has a countably infinite number of parameters, $\boldsymbol\theta = \{\mathbf{w},\boldsymbol\mu,\boldsymbol\sigma\}$: these can be inferred by mean of samples from the underlying distribution $\mathbf{x}$,
\begin{equation}\label{eq:dpgmm}
    p(\boldsymbol\theta|\mathbf{x}) \propto p(\mathbf{x}|\boldsymbol\theta)\,{}p(\boldsymbol\theta) = \prod_j p(x_j|\boldsymbol\theta)\,{}p(\boldsymbol\theta)\,,
\end{equation}
where we assumed that the samples are exchangeable, hence independent and identically distributed. The index $j$ labels the different samples.

In the case of a hierarchical inference, we have access only to the posterior distributions for individual realisations for the target hyper-distribution (as it is the case for GW parameter estimation).
(H)DPGMM models both the hyper-distribution and the individual posterior distributions with DPGMMs, linking them in a hierarchical fashion 
making use of the set of available individual event posterior distributions $\mathbf{Y} = \{\mathbf{y}_1,\ldots,\mathbf{y}_N\}$; hence the name `Hierarchy of Dirichlet process Gaussian mixture models'.
The posterior distribution then reads
\begin{equation}\label{eq:hdpgmm}
    p(\boldsymbol\Theta|\mathbf{Y}) \propto \int \prod_j p(x_j|\boldsymbol\Theta)\,p(x_j|\boldsymbol\theta_j)\,p(\boldsymbol\theta_j|\mathbf{y}_j) \, p(\boldsymbol\Theta) \, \dd \boldsymbol\theta_j \,\dd x_i\,,
\end{equation}
where both $p(x_j|\boldsymbol\Theta)$ and $p(x_j|\boldsymbol\theta_j)$ are DPGMMs. Here, $\boldsymbol\theta_j$ denotes the DPGMM parameters of the $j$-th event and $\boldsymbol\Theta$ denotes the DPGMM parameters $\{\mathbf{w},\boldsymbol\mu,\boldsymbol\sigma\}$ for the hyper-distribution.

The posterior distribution is explored using a variation of the collapsed Gibbs sampling scheme described by \citet{rinaldi:2022:figaro} and implemented in \textsc{figaro}\footnote{\textsc{figaro} is publicly available at \url{https://github.com/sterinaldi/FIGARO}.}, a \textsc{Python} code designed to reconstruct arbitrary probability densities with DPGMM and (H)DPGMM. All the results presented in this paper are obtained making use of this code.

Throughout the paper, for all the calculations that involve cosmological quantities (conversion between luminosity distance and redshift, between detector-frame and source-frame masses and the comoving volume element $\dd V_\mathrm{c}/\dd z$) we will assume the cosmological parameters reported by the Planck collaboration \citep[][Table 1, `Combined' column]{planck:2020}.

\subsection{Selection function}\label{sec:selfunc}
Key ingredient to any astrophysical population inference result is the correction to account for the uneven sensitivity of the detectors to different corners of the parameter space, resulting in different detection probabilities for different events \citep{loredo:2004,mandel:2019}. In particular, the non-parametric approach we are using is even more sensible to the so-called `selection effects', due to the fact that the only source of information about the underlying distribution is the data set on which the inference is based. In this paragraph, we will denote with $\lambda$ the parameters of the GW model: component masses, luminosity distance, spins, etc.

As all the methods based on the Dirichlet process, (H)DPGMM
is only able to reconstruct the observed distribution $\pobs(\lambda|\Lambda)$, rather than the astrophysical one\footnote{Again, we follow the same convention of \citet{rinaldi:2022:hdpgmm}, naming `astrophysical distribution' the one from which the observed events are drawn before accounting for selection effects.} $\pastro(\lambda|\Lambda)$. Here we denote with $\Lambda$ the parameters required by the astrophysical distribution. In this context, the selection effects act as a filter through the so-called `selection function' $\pdet(\lambda)$:
\begin{equation}
    \pobs(\lambda|\Lambda)\propto \pastro(\lambda|\Lambda)\,{}\pdet(\lambda)
    \Rightarrow \pastro(\lambda|\Lambda) \propto \frac{\pobs(\lambda|\Lambda)}{\pdet(\lambda)}\,.
\end{equation}
As pointed out by \citet{essick:2023}, directly inferring the observed distribution $\pobs(\lambda|\Lambda)$ with the astrophysical parameters $\Lambda$ and then correcting for selection effects a posteriori might lead to biased results due to the missed convolution with selection effects. In this framework, we approximate $\pobs(\lambda|\Lambda)$ with (H)DPGMM, thus having an effective representation of this probability density that does not depend on the astrophysical parameters $\Lambda$ but rather on the DPGMM parameters $\boldsymbol\Theta$ introduced in Eq.~\eqref{eq:hdpgmm}:
\begin{equation}\label{eq:selfunc}
    \pastro(\lambda|\Lambda) \propto \frac{\pobs(\lambda|\Lambda)}{\pdet(\lambda)} \simeq \frac{\pobs(\lambda|\boldsymbol\Theta)}{\pdet(\lambda)}\,.
\end{equation}
By doing so, the convolution described in \citet{essick:2023} becomes a simple reweighing that can be done a posteriori in the analysis, thanks to the separation between $\Lambda$ and $\boldsymbol\Theta$.
In other words, being $\Lambda$ the set of parameters of the astrophysical distribution $\pastro(\lambda|\Lambda)$ before the application of selection effects, the inference scheme must account for selection biases directly while exploring the parameter space. Our approach, aiming at describing the observed distribution $\pobs(\lambda|\boldsymbol\Theta)$ only after the application of selection effects, is not directly affected by this bias: in this framework, the parameters $\boldsymbol\Theta$ are not the parameters of the astrophysical distribution, thus the correction can be done a posteriori.

The reconstructed astrophysical distribution is then transformed into a differential merger rate as
\begin{equation}\label{eq:diffmergerrate}
    \frac{\dd^3 \mathcal{R}}{\dd M_1 \dd q \dd z} \propto \frac{\pobs(M_1,q,z|\boldsymbol\Theta)}{\pdet(M_1,q,z)}\,{}(1+z)\,{}\qty(\frac{\dd V_\mathrm{c}}{\dd z})^{-1}\,, 
\end{equation}
where we restricted only to the parameters of interest for this work.

Knowledge of the selection function allows us to correct for the selection bias and recover the astrophysical distribution. It goes without saying that the details of the selection function greatly affect the inference, weighting differently different parts of the parameter space. Due to its crucial role, the greatest care must be put in studying this function.

Traditionally, the selection function is evaluated either via Monte Carlo integration carried out during the population study itself \citep{tiwari:2018,farr:2019,essick:2021} or through (semi-)analytical approximations \citep{wysocki:2019,veske:2021}. Both approaches require a set of injections\footnote{We abide to the LVK nomenclature, indicating simulated BBH signals added to the data as `injections'.} dedicated to studying the performance of the detector in a Monte Carlo fashion, either to be used directly in the analysis (first method) or to calibrate the approximant (second).

In this paper, due to the requirement of exchangeability\footnote{Exchangeability is broken in presence of selection effects: different observations have different weights, inversely proportional to their probability of being observed.}, we cannot include selection effects directly in the inference process: hence, we correct for selection effects a posteriori, after the reconstruction of the observed distribution, as in Eq.~\ref{eq:selfunc}. We rely, therefore, on the analytical approximant for the selection function that will be presented by \citet{lorenzo:2023}\footnote{The code to evaluate our $\pdet$ model and fit its parameters to injection results, and the resulting maximum likelihood parameter values, are publicly available at \url{https://github.com/AnaLorenzoMedina/cbc_pdet}.}. The selection function $\pdet(M_1, M_2, z)$ is modelled as a sigmoid function\footnote{In principle, the selection function depends on the spin parameters as well. Here, we neglect this dependence, assuming that the inclusion of spins in the selection function has a marginal effect on the distribution investigated here: this is based on the fact that the selection function depends weakly on the spin parameters and at the current sensitivity level the precision at which we are able to constrain the spins is low with respect to the other parameters included in this work.}, calibrated on the O3 injections data set released by the LVK collaboration \citep{sensitivityestimate:2021}.
We give further details on this approximant in Appendix~\ref{app:selfunc}.
In the LVK sensitivity estimate campaign, the authors injected events drawn from a fiducial, astrophysically inspired distribution modelled as a non-evolving power-law for both the primary and secondary mass. This is reasonable having in mind the Monte Carlo approach to selection effects: injecting a distribution that is similar in shape to the one that is expected to be found in real data minimises the statistical uncertainty. For the purposes of this work, however, and for non-parametric approaches in general, it is crucial to be able to characterise the detection probability of events that are not predicted by the fiducial model: these are the events that might be smoking guns for the presence of new formation channels. This will be considered in future works.

The way in which we account for selection effects implicitly relies on the assumption that the available data is representative of the actual population, meaning that the underlying distribution has little to no support in regions of the parameter space where either our detector is insensitive or we observed for too little time to have a reasonable chance of detecting events there.
If this assumption fails, the recovered astrophysical distribution will be biased by the fact that the non-parametric reconstruction cannot be informed about the BH population in the so-called `censored' regions of the parameter space and will not be representative of the actual population. We address this potential issue in Appendix~\ref{app:censored}, finding that our results are not likely to be affected by this bias.

As a final note, we want to reiterate that the results of a population study in general, and of a non-parametric one in particular, are extremely dependent on correctly taking into account the corrections for selection effects. 
We believe that the approximant used in this paper captures all the relevant features of the selection function: nonetheless, we cannot exclude the possibility that some of the features henceforth presented could be ascribed to an inaccurate modelling of selection effects. 
To ensure the robustness of our results, we will explore different choices of the selection function as soon as they become available.
Moreover, we are making the assumption that a single selection function calibrated on the O3 injection campaign can be used for a data set composed of observations coming from O1, O2 and O3. This is a reasonable approximation considering that the majority of the GW events has been detected during O3 and the fact that, since we are not reconstructing the total merger rate, we are mainly interested in the shape of the selection function: nonetheless, we plan to investigate the possibility of using different selection functions for different observing runs in a future work.

\subsection{Multivariate plots}
The multivariate plots reporting the differential merger rate density $\dd^3\mathcal{R}/\dd M_1\dd q \dd z$ presented in Sec.~\ref{sec:results} are obtained making use of the specific functional form of the Gaussian mixture model, that allows for an analytical marginalisation and conditioning. In particular, given that the three-dimensional reconstruction obtained with \textsc{figaro} is inherently continuous, our plots do not rely on a redshift binning but simply report slices of the distribution at arbitrary redshift values for visualisation purposes. This does not apply to Figure~\ref{fig:m1-gwtc3}, which is a marginal distribution and thus inherently one-dimensional.

Since (H)DPGMM is meant to reconstruct probability densities, with this specific framework we are not able to estimate the local merger rate density $\mathcal{R}_0$: all the distributions are therefore re-scaled by this quantity and the $y-$axis values are not to be taken as indicative of actual values for the differential merger rate. Lastly, concerning Figs.~\ref{fig:joyplot_q} and~\ref{fig:joyplot_m1}: due to the way in which they are built, each level has its own vertical linear scale, which may vary significantly within the same plot. By construction, every level is re-scaled in such a way that all levels appear to have the same height. We opted for this particular representation because these plots aim at showing the morphological evolution of the differential merger rate rather than than the relative magnitude at different redshifts.

\subsection{BBH sample}
GWTC-3 \citep{GWTC3:2021} adds to the already existing GW catalogues the events detected during the second half of O3, for a total of just below 100 events. Among these observations, \citet{astrodistGWTC3:2023} uses a subset of 62 high-purity BBH events to investigate their astrophysical properties using both astrophysically motivated phenomenological models and semi-parametric models. In this work, we apply the non-parametric method sketched in Section~\ref{sec:methods} to the same subset of events in order to be able to compare our findings to the results presented by the LVK collaboration \citep{astrodistGWTC3:2023}.

\section{Results}\label{sec:results}
\subsection{Primary mass distribution}\label{sec:primarymass}
\begin{figure}
    \centering
    \resizebox{\hsize}{!}{\includegraphics{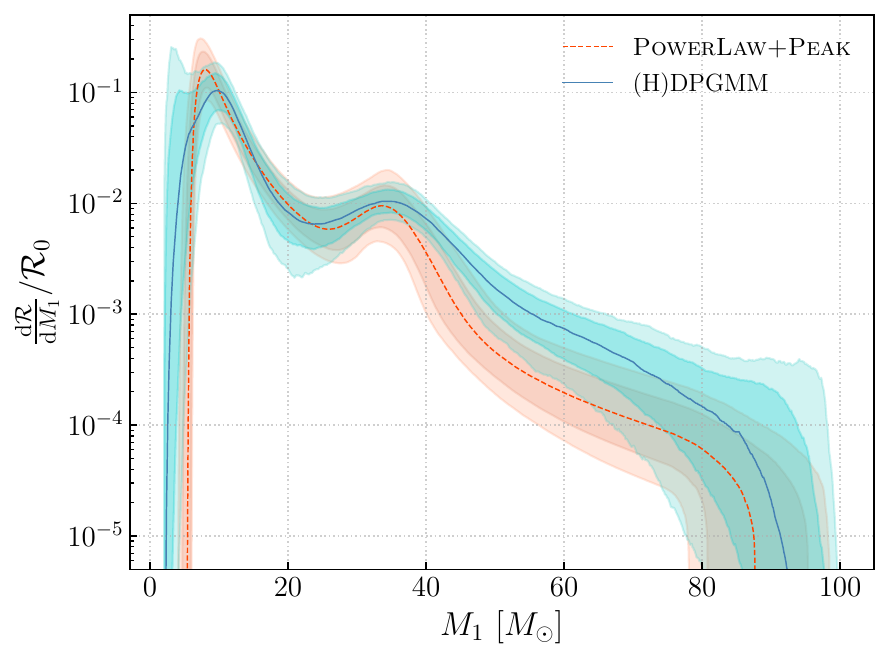}}
    \caption{Primary mass differential merger rate density reconstructed with \textsc{figaro} along with the \plpeak~model from \citet{astrodistGWTC3:2023}. The shaded areas correspond to 68 and 90\% credible regions for (H)DPGMM (blue) and \plpeak~(orange).}
    \label{fig:m1-gwtc3}
\end{figure}

Figure~\ref{fig:m1-gwtc3} shows the marginal primary mass differential merger rate density we obtained with \textsc{figaro}, applied to the subset of 62 high-purity events from GWTC-3 used in \citet{astrodistGWTC3:2023}. We accounted for selection effects using a marginalised version of the three-dimensional selection function approximant.  

Despite being consistent for the majority of the mass spectrum, our non-parametric model departs from the simpler form of the parametric distribution: the power-law behaviour is present only in the low-mass end of the spectrum ($<25\ \msun$) and the smooth tapering of the \plpeak~is less pronounced in our reconstruction.

An excess of BHs is visible at around $35-40\ \msun$, in a region compared with the position of the Gaussian peak of the parametric model \citep{astrodistGWTC2:2021,astrodistGWTC3:2023}. This excess with respect to the simple power-law behaviour, however, is way less localised than the $4.6^{+4.1}_{-2.5}\ \msun$ standard deviation reported in \citet{astrodistGWTC3:2023}, spreading all the way up to $\sim 70\ \msun$. 
Lastly, the non-parametric distribution shows a cutoff at around $90\ \msun$ (mainly due to GW190521, \citealt{GW190521:discovery}).

We made sure that the features we observe, especially the excess between $35$ and $70\ \msun$, are not an artefact of the selection function repeating the analysis with a marginalised version of the two-dimensional (primary and secondary mass) approximant presented in \citet{veske:2021}. We find that the features are robust in this respect.

From the mass distribution alone it is understandably difficult to decide whether the features we find in the BH mass distribution presented in this section are intrinsic or due to stochastic fluctuations, let alone providing a reliable interpretation in terms of astrophysical processes.
We therefore exploited the possibility offered by (H)DPGMM of inferring multivariate probability density to investigate the correlations among the primary mass $M_1$, the mass ratio $q$ and the redshift $z$: the presence of such correlations will both confirm the astrophysical origin of the features and guide the development of theoretical models for BH formation channels.

\subsection{Joint primary mass, mass ratio, and redshift distribution}\label{sec:m1qz}
In this section, we report and describe the non-parametric reconstruction of the differential merger rate as obtained with \textsc{figaro}.

\begin{figure}
    \centering
    \resizebox{\hsize}{!}{\includegraphics{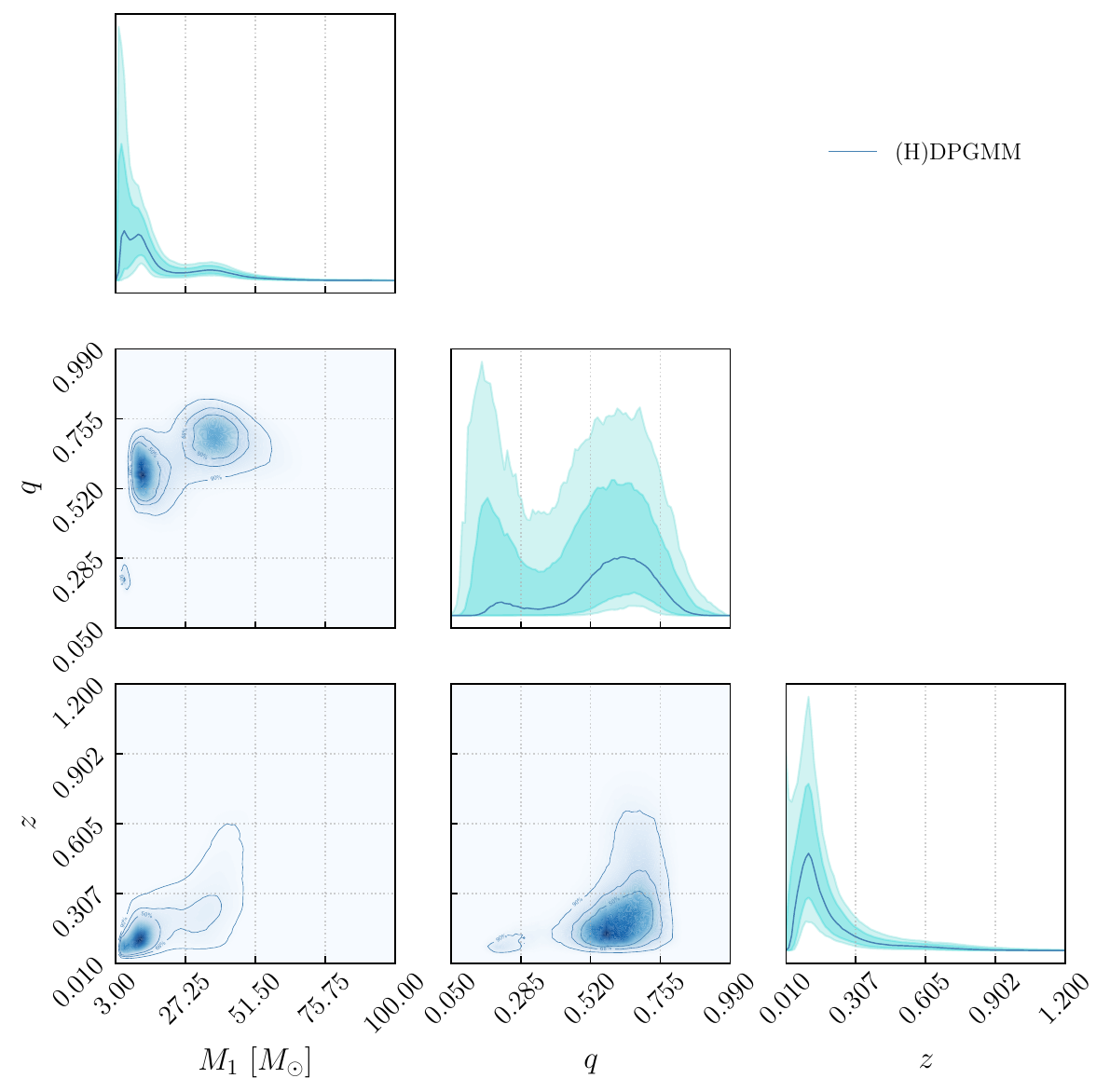}}
    \caption{$M_1$, $q$ and $z$ differential merger rate density reconstructed with \textsc{figaro} as in Eq.~\ref{eq:diffmergerrate}. The diagonal panels show the marginal distributions for $M_1$, $q$ and $z$ (top to bottom), respectively, and the panels below diagonal show the median distribution marginalised over the third variable.}
    \label{fig:astro_m1qz_bbh_rate}
\end{figure}

The three-dimensional distribution displayed as a corner plot in Fig.~\ref{fig:astro_m1qz_bbh_rate} shows correlations among all possible pairs of parameters. In particular, we find worth mentioning the following two things:
\begin{itemize}
    \item \textbf{The mass ratio does not show support for symmetric binaries:} this is particularly interesting since most binary evolution models predict the mass ratio to have a maximum at $q\sim 1$ \citep{dominik:2012,belczynski:2016,demink:2016,giacobbo:2018,klencki:2018,broekgaarden:2022}. The central panel of Fig.~\ref{fig:astro_m1qz_bbh_rate} shows a bimodal marginal distribution for the mass ratio, with the main mode centred around $q=0.7$. This particular feature is not an artefact of either the model or the code we are employing, as shown in Appendix~\ref{app:validation}.
    \item \textbf{The mass distribution evolves with redshift:} the lower-left panel of Fig.~\ref{fig:astro_m1qz_bbh_rate} shows that the primary mass changes with redshift: at high redshift, the mass distribution shows no support for BHs with masses $<20\ \msun$. Conversely, for low $z$, the distribution peaks at around $8-10\ \msun$ with a cutoff just above $40\ \msun$.
    This is in contrast with the findings of \citet{astrodistGWTC2:2021} and \citet{ray:2023}, who do not find evidence for the evolution of the mass distribution. \citet{karathanasis:2023}, on the other hand, find an indication of such redshift dependence.
\end{itemize}

In the following, we will describe the different features that are present in the evolving primary mass and mass ratio distribution, deferring a discussion of such in terms of astrophysical models to Section~\ref{sec:interpretation}.

\subsubsection{Mass ratio}
\begin{figure*}
    \centering
    \subfigure[]{
        \includegraphics[width = 0.99\columnwidth]{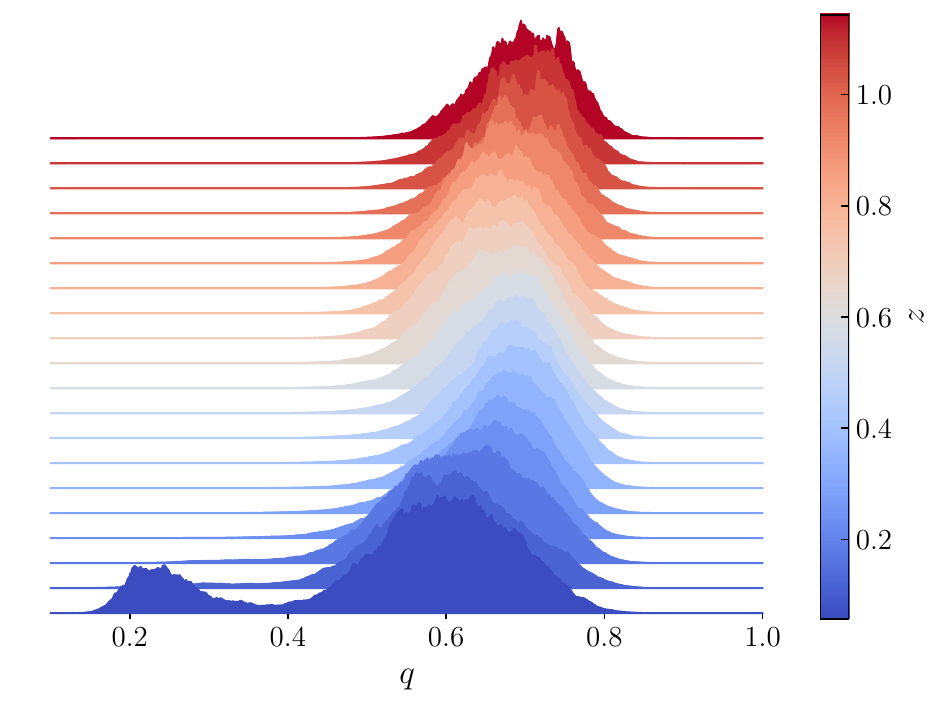}\label{fig:joyplot_q}
        }
    \subfigure[]{
        \includegraphics[width = 0.99\columnwidth]{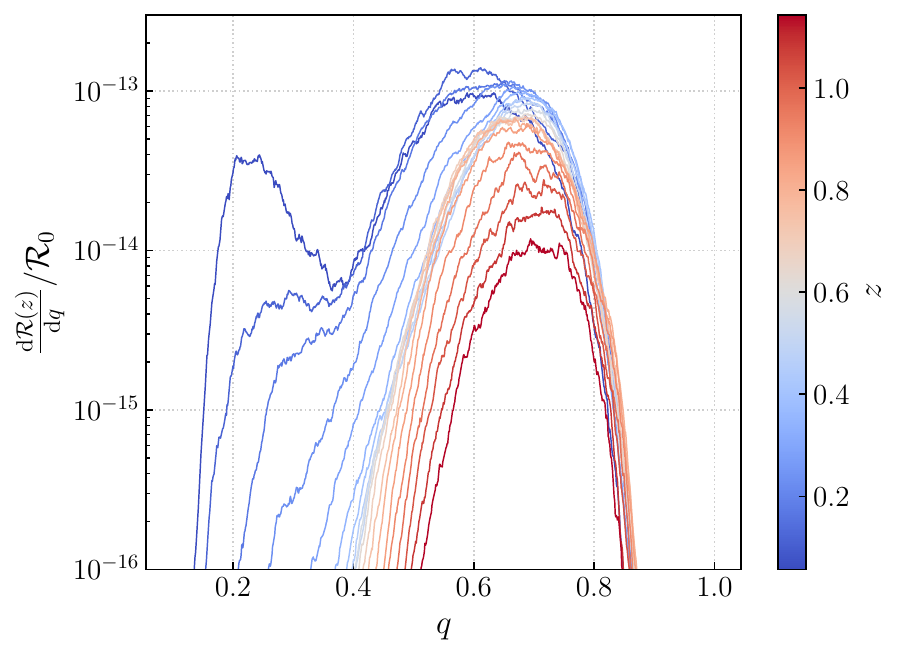}\label{fig:qvsz_cond_logscale}
        }
    \caption{Mass ratio evolution with redshift: joy plot (a) and tomography (b). Both panels report the median distribution only: an animated version of this plot including the 68\% credible regions is available as online-only supplementary material along with plots showing individual slices. This distribution, set aside an excess at low redshifts ($z\lesssim$ 0.2), is consistent throughout the whole redshift spectrum in having support at around $0.7$, with negligible support for $q = 1$.}\label{fig:q_dist}
\end{figure*}

Figure~\ref{fig:q_dist} shows the evolution with redshift of the mass ratio differential merger rate, $\dd \mathcal{R}/\dd q$. Overall, the mass ratio distribution is roughly constant at all redshifts, being composed of a single pileup at $q \sim 0.7$. The only additional feature is a peak at around $q\sim 0.25-0.3$ at redshift $z\lesssim0.2$, populated by two events only (GW190412 and GW190917\textunderscore114630).
This latter feature could simply be due to the small number of events: with more events coming from the current and future observing run, it will be possible to assess its actual significance in the astrophysical distribution.

The differential merger rate in Fig.~\ref{fig:qvsz_cond_logscale} is not precisely Gaussian: in particular, the distribution favours asymmetric binaries, with a preference for smaller redshifts. Symmetric binaries, conversely, are equally disfavoured independently of the redshift, being suppressed just above $q = 0.9$ at 90\% credible level (see Figure~\ref{fig:astro_m1qz_bbh_rate}, second row, second column panel and online-only material): support for symmetric binaries is practically absent. 
A more detailed discussion of the shape of this distribution in terms of observed events can be found in Appendix~\ref{app:qdist}.

This is in contrast with the parametric model used in \citet{astrodistGWTC3:2023}, where the mass ratio is modelled as a power-law: $\dd \mathcal{R}/\dd q \propto q^{\beta}$, with $\beta = 1.1^{+1.7}_{-1.3}$. The specific parametric model employed in this analysis inherently favours symmetric binaries, having a maximum at $q = 1$ for $\beta > 0$, potentially making the preference for symmetric systems an `artefact' of the model. 
Other works making use of semi-parametric models investigate this distribution \citep{tiwari:2022,callister:2023}, not finding the results presented in this work: \citet{tiwari:2022}, however, makes use of a model based on a power-law for $q$ \citep{tiwari:2021:vamana}.
\citet{edelman:2023} hint at the possibility of a decrease in the merger rate near equal mass binaries.
The suppression of symmetric binaries, not predicted by the majority of the existing literature on the subject, is however confidently found in the data by our non-parametric model. 

In order to assess that this feature is not an artefact induced by the model we are using, we validated our method using a simulated data set modelled after the findings of \citet{astrodistGWTC3:2023}. The results, presented in Appendix~\ref{app:validation}, show that our model is capable of retrieving the simulated distribution, including a power-law $q^{1.1}$ for the mass ratio.

\subsubsection{Primary mass}
\begin{figure*}
    \centering
    \subfigure[]{
        \includegraphics[width = 0.99\columnwidth]{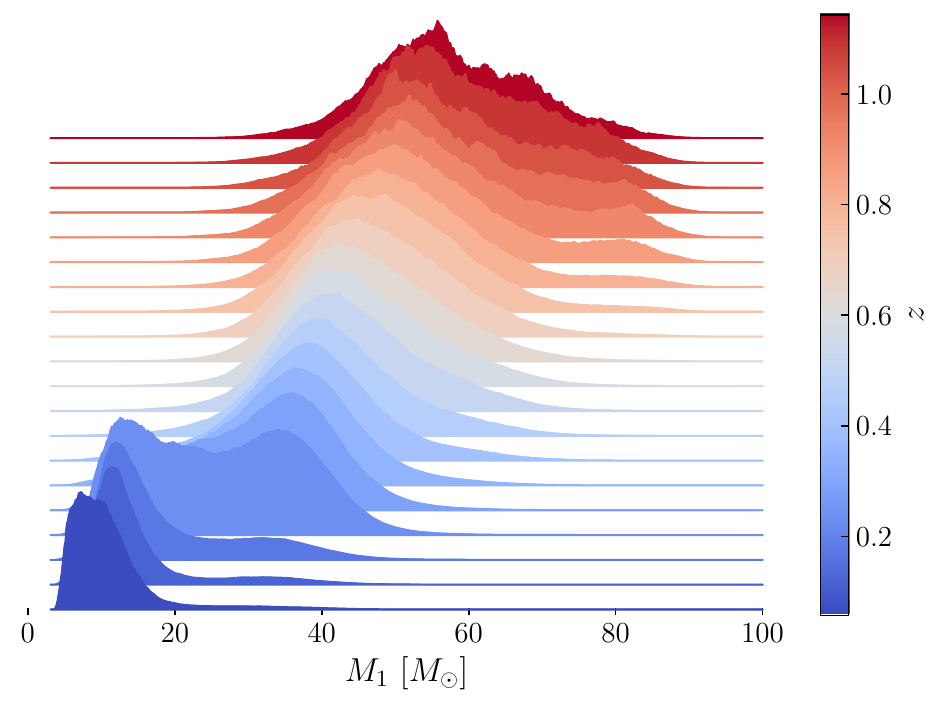}\label{fig:joyplot_m1}
        }
    \subfigure[]{
        \includegraphics[width = 0.99\columnwidth]{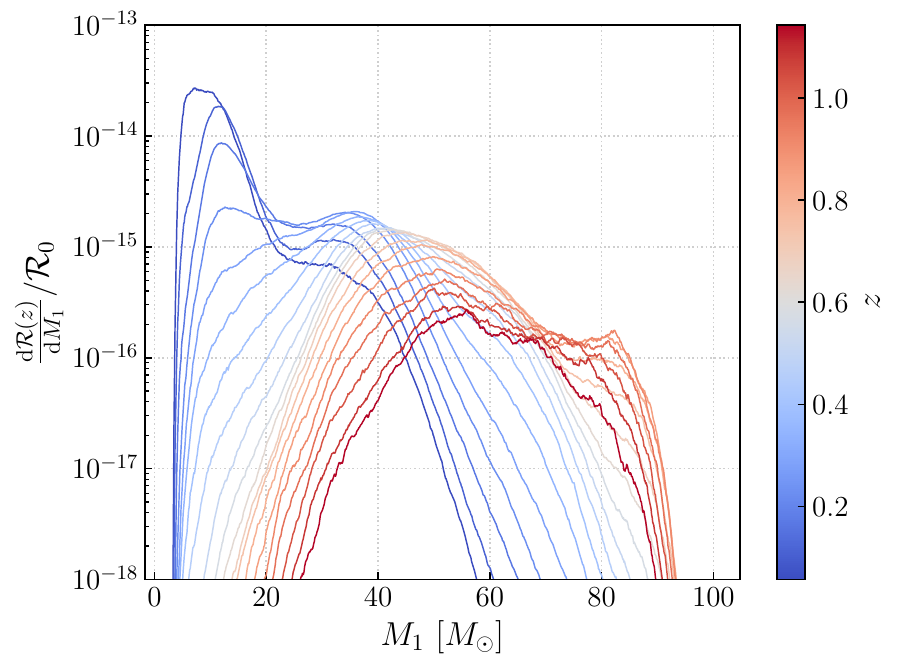}\label{fig:M1vsz_cond_logscale}
        }
    \caption{Primary mass evolution with redshift: joy plot (a) and tomography (b). Both panels report the median distribution only: an animated version of this plot including the 68\% credible regions is available as online-only supplementary material along with plots showing individual slices. The mass distribution drastically changes with redshift, suggesting the presence of two distinct populations of BHs.}\label{fig:M1_dist}
\end{figure*}

The primary mass differential merger rate is reported in Fig.~\ref{fig:M1_dist}. The primary mass distribution changes its morphology with redshift, from a power-law-like distribution with a preference for low-mass objects at redshift $z < 0.4$ to a high-mass pileup for $z > 0.4$. 
The combination of these two sub-populations, once marginalised over redshift, is the basis for the \plpeak~model.

\citet{astrodistGWTC3:2023} find no evidence in favour of a redshift dependence of the primary mass, in contrast with our findings. We ascribe this difference to the  parametric model \citet{fishbach:2021,astrodistGWTC3:2023} adopted to investigate this correlation: they assume that the parameters of the mass distribution change with redshift, rather than allowing for a change in the functional form. In particular, they model the high-mass tail of the distribution with a separate power-law index with the  cutoff location evolving with redshift. 
Based on our findings, this model does not properly capture the actual evolution of the mass spectrum. 
We believe that this is the main reason why \citet{astrodistGWTC3:2023} do not find evidence in favour of the evolution of the BH mass function with redshift: while performing model selection in the context of Bayesian statistics, the addition of one or more degrees of complexity (such as the second power-law index) that do not improve the modelling of the available data is penalised by the Bayes' factor.

Given our non-parametric reconstruction, we believe that a good phenomenological parametric model, still based on the \plpeak~model, would be a superposition of a power-law and a Gaussian peak with a redshift-dependent relative weight $w(z)$, to smoothly transition from the low-redshift, low-mass power-law to the high-redshift, high-mass Gaussian peak with a mean that evolves with redshift.
Using a model along the lines of this one, a parametric analysis should in principle be able to pick up the redshift evolution of the mass function.
\citet{vanson:2022} propose a similar model where the relative weights of the power-law and the Gaussian peak are proportional to $(1+z)^{\kappa_\mathrm{pl}}$ and $(1+z)^{\kappa_\mathrm{peak}}$ respectively. The authors find marginal evidence in favour of the evolution of the mass with redshift.

\subsubsection{Transition from power-law to peak}
\begin{figure}
    \centering
    \resizebox{\hsize}{!}{\includegraphics{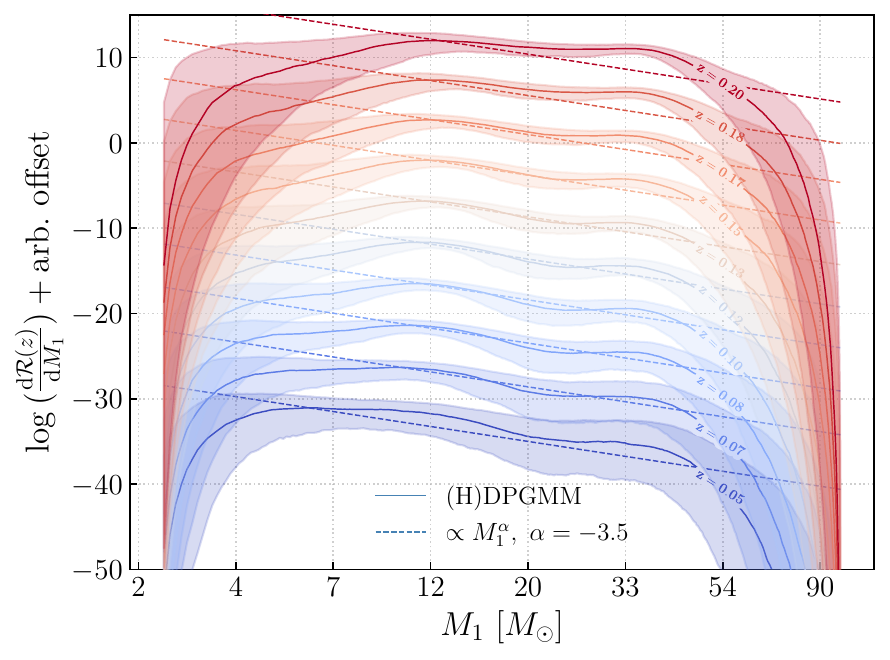}}
    \caption{Primary mass conditional differential merger rate density} for the local Universe ($z\leq0.2$) compared with a power-law distribution (index $\alpha$ from \citealt{astrodistGWTC3:2023}). Shaded areas correspond to 68\% credible regions.
    \label{fig:M1vsz_pl}
\end{figure}

Figure~\ref{fig:M1vsz_pl} compares the mass distribution in the local Universe, $z \leq 0.2$, with a pure power-law distribution with index $\alpha = -3.5$, as found in \citet{astrodistGWTC3:2023}. This definition of local Universe is purely arbitrary and chosen for visualisation purposes.
The non-parametric reconstruction is consistent with the power-law distribution in the $10\sim40\ \msun$ region, up to redshift $z \sim 0.13$, with cutoffs at both ends of the mass spectrum. 
At larger redshift, the distribution starts to deviate from the simple power-law behaviour, transitioning towards the $\sim 35\ \msun$ Gaussian-like feature. Given this qualitative agreement between our reconstruction and the power-law reported by \citet{astrodistGWTC3:2023}, we suggest that the \plpeak~model is actually driven by the local merger rate density, with a contribution for the Gaussian peak by the BHs further away in look-back time.

\subsubsection{Peak evolution with redshift}
\begin{figure}
    \centering
    \resizebox{\hsize}{!}{\includegraphics{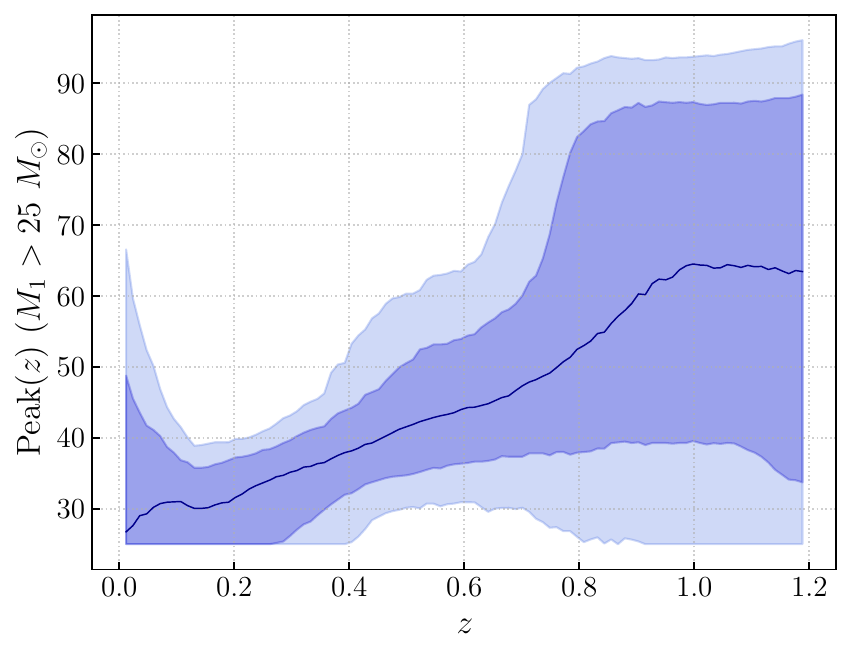}}
    \caption{Evolution of the peak of the pileup feature of the mass distribution with redshift. The peak is defined as the maximum of the distribution for $M_1 > 25\ \msun$. The shaded areas correspond to the 68 and 90\% credible regions.}
    \label{fig:peak_evolution}
\end{figure}

Figure~\ref{fig:joyplot_m1} shows that the pileup drifts towards larger masses with redshift. This behaviour is highlighted in Fig.~\ref{fig:peak_evolution}. We define the peak of the pileup as the maximum of the distribution for\footnote{This threshold between low-mass and high-mass regime is purely heuristic and not backed by any specific model or figure of merit. We defer a more detailed analysis of this threshold to a future work.} $M_1 > 25\ \msun$.
With this, we can isolate three different regimes:
\begin{itemize}
    \item $z < 0.2$: this region is dominated by the low-mass power-law, therefore the maximum leans towards $25\ \msun$, driven by the preference of the local Universe for lighter BHs. The pileup here, if present, is subdominant; 
    \item $0.2 < z < 0.7$: in this redshift interval the number of available events is large enough to have an accurate reconstruction of the pileup: its peak steadily moves towards larger masses with look-back time, with a somewhat linear dependence;
    \item $z > 0.7$: the number of events beyond $z = 0.7$ is too limited: this fact, combined with a poor resolution of the binary parameters, makes the non-parametric reconstruction merely a hint of the actual behaviour of the underlying distribution. Therefore, the peak in this region is ill-defined, and its 90\% credible interval encompasses almost all the mass spectrum.
\end{itemize}

\subsubsection{Minimum and maximum mass of a black hole}
Following \citet{astrodistGWTC3:2023}, we define the minimum and maximum mass of a BH as the 1st and 99th percentile of the mass distribution, $M_{1\%}$ and $M_{99\%}$. We report these values in Table~\ref{tab:min_max_mass}. The redshift values are the same used for the primary mass joy plot. For the same quantities, \citet{astrodistGWTC3:2023} report $M_{1\%} = 5.0^{+0.9}_{-1.7}$ and $M_{99\%} = 44^{+9.2}_{-5.1}$: both values are in agreement with our findings for the low-redshift region, in which the mass distribution is dominated by the power-law-like feature. 

We did not quote the credible interval for $M_{1\%}$ for redshifts above $0.25$ due to the dominance of selection effects in the low-mass, high-redshift region of the parameter space. The vetted region bounds below the credible interval for this quantity: we deemed the uncertainty estimate for $z > 0.25$ unreliable, thus opting for quoting the median value only, for reference purposes and to show that a growing trend is however present.

\begin{table}
    \centering
    \caption{Minimum and maximum value for the BH mass as a function of redshift. For the minimum mass, we quote the credible interval only for $z<0.25$ due to the dominance of selection effects in the low-mass, high-redshift region of the parameter space.}
    \begin{tabular}{lll}
    $z$ & $M_{1\%}\ [\msun]$ & $M_{99\%}\ [\msun]$ \\
    \midrule
    0.057 & $4.6^{+1.6}_{-1.0}$ & $36^{+12}_{-18}$ \\
    0.114 & $5.5^{+2.2}_{-1.6}$ & $41^{+8}_{-7}$ \\
    0.171 & $6.5^{+2.6}_{-2.6}$ & $45^{+8}_{-6}$ \\
    0.228 & $7.4^{+3.7}_{-3.5}$ & $49^{+10}_{-6}$ \\
    0.286 & $8.0$ & $54^{+12}_{-7}$ \\
    0.343 & $9.3$ & $60^{+11}_{-10}$ \\
    0.400 & $12$ & $64^{+11}_{-10}$ \\
    0.457 & $15$ & $67^{+11}_{-10}$ \\
    0.514 & $18$ & $70^{+13}_{-10}$ \\
    0.571 & $21$ & $73^{+15}_{-10}$ \\
    0.628 & $24$ & $77^{+14}_{-12}$ \\
    0.685 & $25$ & $80^{+12}_{-14}$ \\
    0.742 & $26$ & $84^{+9}_{-15}$ \\
    0.800 & $28$ & $85^{+9}_{-15}$ \\
    0.857 & $29$ & $85^{+9}_{-15}$ \\
    0.914 & $31$ & $85^{+9}_{-16}$ \\
    0.971 & $33$ & $84^{+10}_{-17}$ \\
    1.029 & $35$ & $83^{+11}_{-20}$ \\
    1.086 & $37$ & $81^{+12}_{-21}$ \\
    1.143 & $38$ & $80^{+14}_{-25}$ \\
    \bottomrule
    \end{tabular}
    \label{tab:min_max_mass}
\end{table}

\subsubsection{GW190521: a stand-alone?}
\begin{figure}
    \centering
    \resizebox{\hsize}{!}{\includegraphics{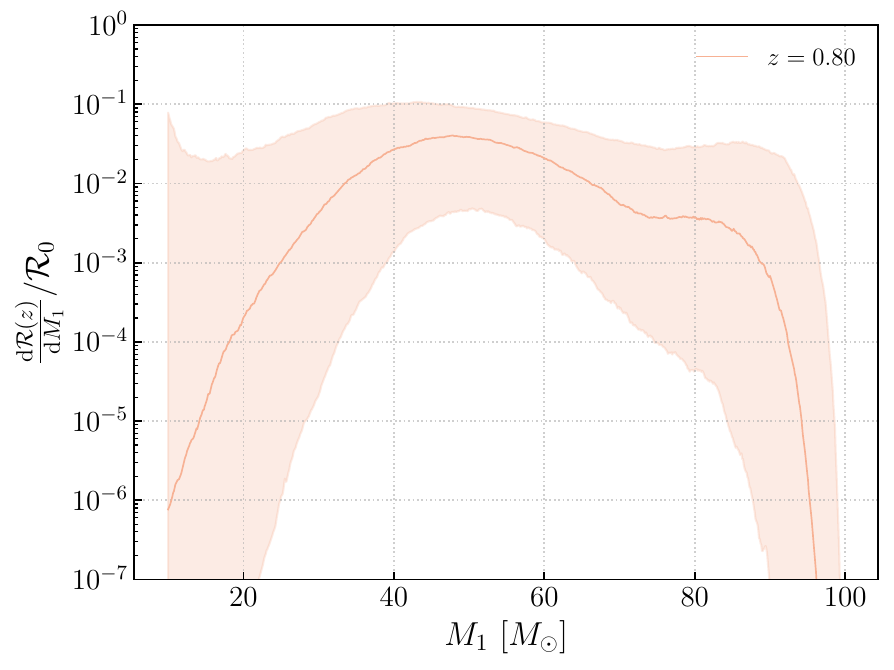}}
    \caption{Primary mass conditional differential merger rate density} for $z = 0.8$. The shaded area correspond to the 68\% credible region. The peak at $M_1\sim 90\ \msun$ marks the presence of GW190521.
    \label{fig:GW190521_redshift_slice}
\end{figure}

The primary mass spectrum also shows a small peak\footnote{We mention, however, that this additional feature is not as robust as the $35-50\ \msun$ one due to the large uncertainties.} at $M_1\sim 90\ \msun$, $z\sim 0.8$ (see Figure~\ref{fig:GW190521_redshift_slice}). 

This feature marks the presence of GW190521 \citep{GW190521:discovery}, the most massive BBH in the high-purity catalogue we are considering here, with a total mass of $\sim 150\ \msun$. 
In \citet{GW190521:implications} the primary mass of this BBH is said to be in tension with the population inferred from O1 and O2, thus suggesting the possibility for this event to be an `outlier'. Despite its peculiar component masses, in the latest population studies (most notably \citealt{astrodistGWTC2:2021} and \citealt{astrodistGWTC3:2023}) GW190521 is found to be consistent with the rest of the BBH population. Here we do not quantitatively assess whether if this event is an outlier or not, but rather we take the occasion to briefly discuss the concept of outliers in the context of non-parametric methods.

An outlier is often defined as a data point that is difficult to explain with or is unlikely to have been produced by a specific population model: this is the case, for example, of a point that has been drawn by a sub-population not included in the model. However, in the context of non-parametric analysis we cannot really talk about outliers in this sense, given that the inferred distribution is the one that best describes \emph{all} the available data, regardless of their origin. It is up to the analysts to decide whether to include a certain data point (or, in our case, an event) in the data set. An event or a group of events that shows different properties from the rest of the catalogue, in a non-parametric analysis, is simply modelled as a sub-population on its own, as it is also the case for the mass ratio of GW190412 and GW190917\textunderscore114630. We therefore think that, in this context, `stand-alones' could be a better name for these events.

GW190521 could be, qualitatively, a stand-alone event in this sense, and might be hinting towards the presence of a new, previously unobserved sub-population of BBHs. This possibility is supported by the results presented by \citet{morton:2023}, where we find evidence for GW190521 to be formed in an AGN  disk. 
In AGN disks, BBHs might form with masses that reach and exceed $100\ \msun$, due to the increased probability for hierarchical mergers and accretion \citep{mckernan:2012,mckernan:2018,bellovary:2016,bartos:2017,stone:2017,yang:2019,tagawa:2020,gayathri:2022}. GW190521 could be the first observed BBH to belong to this sub-population \citep{graham:2020}.

\section{Discussion: interpretation of the astrophysical distribution}\label{sec:interpretation}
We now discuss the astrophysical implications of the distribution reported in the previous section. In what follows, we will assume that the inferred distribution is correct and not affected by the possible caveats discussed above.
\subsection{Primary mass}
We find that the astrophysical population of BBHs changes with look-back time. 
Our main result is the morphological evolution of the distribution of primary masses as a function of redshift. In the low-redshift Universe most BBHs have primary mass $<20\ \msun$, in agreement with the X-ray binaries in the Milky Way \citep[e.g.,][]{ozel:2010,farr:2011}. In contrast, at redshift $z\geq0.4$, the dominant population has primary mass $\geq20\ \msun$.

This evolution of the primary mass with redshift has several possible, non-mutually exclusive interpretations. First, we can think of a metallicity trend: in the low-redshift Universe, we  expect to observe mostly BBHs born from metal-rich progenitors, whereas in the high redshift Universe most BBHs should originate from metal-poor ($Z\leq10^{-4}$) and metal-free progenitors  \citep{hartwig:2016,kinugawa:2016,inayoshi:2017,mapelli:2017,schneider:2017,tanikawa:2021,costa:2023}. The primary mass of BBHs born from metal-rich stars is expected to peak at $\sim{8-12}\ \msun$, because of efficient wind mass loss   \citep{belczynski:2016,stevenson:2017,giacobbo:2018,klencki:2018,kruckow:2018,spera:2019,schneider:2021,iorio:2023,fragos:2023,agrawal:2023}. Conversely, the low-mass peak of the primary mass function is suppressed in BBHs born from metal-poor and metal-free stars (see, e.g., Figure~8 of \citealt{santoliquido:2023}): most of them have primary mass in the $\sim{20-50}\ \msun$ range, possibly enforcing the peak at $\sim{35}\ \msun$ \citep{kinugawa:2014,kinugawa:2020,liubromm:2020,tanikawa:2021GW190521,vink:2021,tanikawa:2022,costa:2023,santoliquido:2023}. However, current models do not predict that the low-mass peak disappears already at $z\sim 0.4$, but rather at higher redshift ($z\geq6$, \citealt{santoliquido:2023}). 

A second possible interpretation for the evolution of the primary mass function is a change of the relative importance of different formation channels with redshift. For example, the dominant formation channel might change from isolated binary evolution to dynamical assembly, as redshift increases. Several previous studies indicate that we expect higher BBH masses from dynamical formation  in dense star clusters \citep[e.g.,][]{rodriguez:2016,askar:2017,chatterjee:2017,antonini:2019,kremer:2020,torniamenti:2022,antonini:2023} and AGN disks \citep[e.g.,][]{mckernan:2012,mckernan:2018,bellovary:2016,bartos:2017,stone:2017,tagawa:2020}. In particular, globular clusters' formation peaks at redshift $z\sim 3-4$ \citep{vandenberg:2013,choksi:2018,elbadry:2019}. This redshift-dependent formation history, combined with longer-than-expected delay times and/or with a higher-than-expected merger efficiency, might produce a dominant population of massive BH mergers down to a redshift $z\sim 0.4$ \citep{rodriguez:2018,choksi:2019,antonini:2020,antonini:2023,mapelli:2022}.

If this interpretation is correct, we should observe an evolution with redshift not only of the primary mass function but also of the effective and precessing spin parameters. In fact, dynamical interactions randomise the orientation of BH spins, leading to a nearly-isotropic spin distribution, while isolated binary evolution tends to align BH spins with the orbital angular momentum of the binary systems \citep[e.g.,][]{kalogera:2000,rodriguez:2016spin}.
Moreover, hierarchical BBH mergers are associated with large spin magnitudes, at least for the primary BH, because the dimensionless spins of merger remnants peak at $\sim{0.7}$ \citep{buonanno2008,jimenez:2017}. Hence, an increase of the hierarchical merger population with redshift would result into a broadening of the effective spin distribution with redshift, and would lead to the emergence of a peak in the precessing spin around $\sim{0.7}$ \citep{baibhav:2020,mapelli:2021}. 
The selection function implementation used in this work does not account for spins, therefore we did not include these parameters in this work. We plan to investigate the effective spin distribution in a future paper. 
Here, we mention that \cite{biscoveanu:2022} find an
increase in the width of the effective spin distribution of BBHs with increasing redshift. This might be a hint of a larger contribution of the hierarchical merger scenario at higher redshift \citep{baibhav:2020,mapelli:2021, kritos:2022,santini:2023}, but it is still consistent with other scenarios. 

A third hypothesis (not necessarily alternative to the other two) is that the evolution of the primary mass function captured by our method stems from processes that current models fail to properly describe. Overall, models that do not predict any evolution (or predict only a mild evolution) with redshift \citep[e.g.,][]{mapelli:2019} are already challenged by our findings.

\subsection{Mass ratio}
Our results  seem to indicate that the preferred BBH mass ratio is $q \sim 0.7$, with little or no evolution with redshift (with the exception of the lowest redshift region, dominated by the events GW190412 and GW190917\textunderscore114630). In contrast, most current models (especially isolated binary evolution via common envelope and chemically homogeneous evolution) indicate a preference for equal-mass or nearly-equal mass systems \citep[e.g.,][]{belczynski:2016,demink:2016,giacobbo:2018,klencki:2018,kruckow:2018,broekgaarden:2022}, with a few exceptions \citep{vanson:2022,santoliquido:2023}.

Most dynamical formation channels allow for a larger fraction of unequal-mass systems than isolated binary evolution \citep[e.g.,][]{dicarlo:2019,arcasedda:2020,torniamenti:2022}. In particular, BBHs born from hierarchical mergers have a strong preference for unequal mass mergers \citep{rodriguez:2019,fragione:2020,zevin:2022}. However, recent studies show that hierarchical mergers cannot account for the majority of LVK events \citep{zevin:2022,fishbach:2022}.

Recently, \cite{santoliquido:2023} have shown that metal-poor and metal-free BBH mergers tend to have $q\sim{0.7}$ at redshift $z\lesssim 5$ (see their Fig.~7). This feature is mostly an effect of the evolutionary channel: in their models, all the BBHs with $q\leq 0.7$ form from binary systems in which the progenitor star of the primary BH retains at least a fraction of its  H-rich envelope at the time it collapses to BH. In most cases, the binary system evolves via stable mass transfer. Similarly, \cite{vanson:2022} find that BBHs formed via stable mass transfer in binary systems peak at $q\sim 0.6-0.7$ (see their Fig.~9). Notably, the revised treatment of Roche-lobe overflow stability by \cite{olejak:2021} leads to significantly lower values of $q\sim{0.4-0.6}$ (see their Figure 4): most of the unequal-mass BBHs in their model form without common-envelope evolution. Clues in this direction are also reported in \citet{gallegosgarcia:2021}.

Overall, none of the current models in the literature is able to explain at the same time a non-evolving mass ratio $q\sim 0.7$ and the rapid evolution of the primary mass with redshift we find with (H)DPGMM. This might indicate that astrophysical models fail to describe some key physical process, for example the efficiency of mass accretion and angular momentum transport during mass transfer \citep[e.g.,][]{gallegosgarcia:2023,willcox2023}, the impact of stellar rotation \citep[e.g.,][]{mapelli:2020,marchant:2020,riley:2021}, or the efficiency of dynamical assembly \citep[e.g.,][]{fragione:2018,mapelli:2022}.

\section{Conclusions}\label{sec:conclusions}
We have applied our non-parametric model, (H)DPGMM, to the GW events included in GWTC-3, focusing on the primary mass, mass ratio, and redshift distribution.
Such distribution, once corrected for selection effects, clearly shows the presence of correlation among the three parameters, providing support in favour of the evolution of the primary mass distribution with redshift. In particular, we found that:
\begin{itemize}
    \item The mass ratio distribution does not show support for symmetric binaries, favouring systems with $q\sim 0.7$ mostly independent of redshift;
    \item The primary mass distribution is composed of two distinct sub-populations: a low-mass, low-redshift power-law-like population and an high-mass, high-redshift Gaussian-like feature. The change between the two sub-populations happens between $z\sim 0.2$ and $z\sim 0.4$;
    \item The position of the Gaussian-like feature evolves with redshift, drifting towards larger masses with look-back time.
\end{itemize}

We outlined two possible, non-mutually exclusive explanations for the evolution of the astrophysical BBH population with look-back time. The mass could follow a trend in metallicity, changing from low-mass BHs born from metal-rich progenitor stars at low redshift to more massive BHs, produced by metal-poor and metal-rich stars. The other possible interpretation is a change of the relative importance of different formation channels, from a isolated evolution-dominated scenario at $z < 0.4$ to a prevalent contribution of dynamically assembled systems for $z > 0.4$.

Our findings for the mass ratio are in contrast with most of the current models for isolated binary evolution, where the favoured systems are symmetric or near-symmetric binaries; unequal-mass systems are more likely to be produced by dynamical formation channels or hierarchical mergers. The apparent contradiction between the rapid evolution of the BH mass function and the constancy of the mass ratio distribution could hint towards some missing physics in current astrophysical models.

The results presented in this paper will be crucial not only for the advancement of BH astrophysics, but also for all those investigations that, at different levels, rely on the BH mass distribution. For example, the mass function is a key ingredient of methods for the inference of cosmological parameters \citep{cosmoO3:2023}, the galaxy catalogue method \citep{gray:2020}, and the joint population and cosmology approach \citep{mastrogiovanni:2021}. 
Moreover, the presence of non-scale-invariant and redshift-evolving features like the pileup at $z > 0.4$ will have a dramatic impact on the methods that relies on features in the mass distribution to infer the cosmological parameters \citep{farr:2019:cosmology,mancarella:2022,mukherjee:2022}.

As a final remark, we remind that the results presented in this paper are obtained under two important caveats, namely that the data are representative of the underlying population (we are not in presence of censored data and the features in the underlying distribution are properly sampled) and that the approximant we employed captures all the features of the selection function. We made sure that these assumptions are met (see Appendices~\ref{app:selfunc},~\ref{app:validation} and~\ref{app:censored}).  Nonetheless, it is still possible that some of the features we reported are to be ascribed to the limited number of available GW events: the new events detected during the currently ongoing LVK run, together with the flexibility of data-driven models such as (H)DPGMM to adapt themselves to new observations, will either confirm or disprove the findings reported in this paper and will unveil new features of the BH population.

\begin{acknowledgements}
The authors are grateful to Christopher~Moore, Thomas Callister and the anonymous referee for useful comments. We are also thankful to the organisers and participants of GWPopNext (University of of Milano-Bicocca, July 2023) for stimulating discussion.
MM acknowledges financial support from the European Research Council for the ERC Consolidator grant DEMOBLACK, under contract no. 770017, and from the German Excellence Strategy via the Heidelberg Cluster of Excellence (EXC 2181 - 390900948) STRUCTURES.
This work has received financial support from Xunta de Galicia (CIGUS Network of research centers) and by European Union ERDF.  ALM and TD are supported by research grant PID2020-118635GB-I00 from the Spanish Ministerio de Ciencia e Innovaci{\'o}n. 

This research has made use of data or software obtained from the Gravitational Wave Open Science Center (gwosc.org), a service of the LIGO Scientific Collaboration, the Virgo Collaboration, and KAGRA. This material is based upon work supported by NSF's LIGO Laboratory which is a major facility fully funded by the National Science Foundation, as well as the Science and Technology Facilities Council (STFC) of the United Kingdom, the Max-Planck-Society (MPS), and the State of Niedersachsen/Germany for support of the construction of Advanced LIGO and construction and operation of the GEO600 detector. Additional support for Advanced LIGO was provided by the Australian Research Council. Virgo is funded, through the European Gravitational Observatory (EGO), by the French Centre National de Recherche Scientifique (CNRS), the Italian Istituto Nazionale di Fisica Nucleare (INFN) and the Dutch Nikhef, with contributions by institutions from Belgium, Germany, Greece, Hungary, Ireland, Japan, Monaco, Poland, Portugal, Spain. KAGRA is supported by Ministry of Education, Culture, Sports, Science and Technology (MEXT), Japan Society for the Promotion of Science (JSPS) in Japan; National Research Foundation (NRF) and Ministry of Science and ICT (MSIT) in Korea; Academia Sinica (AS) and National Science and Technology Council (NSTC) in Taiwan.
\end{acknowledgements}

\bibliographystyle{aa}
\bibliography{bibliography.bib}
\begin{appendix}
\section{Selection function approximant}\label{app:selfunc}
Inference from the catalogue of GW events detected by LVK \citep{GWTC3:2021} is subject to observation selection effects \citep[e.g.][]{mandel:2018}, as the great majority of BBH mergers go undetected and only signals `loud' enough to cause significant triggers in dedicated searches are followed up with detailed parameter estimation. Here, as in the most recent LVK population analysis \citep{astrodistGWTC3:2023}, only candidates with an estimated false alarm rate (FAR) of 1 per year or less in one or more searches are considered.
We refer to such signals as `detected' or `found'. To enable accurate estimates of the selection function, a large-scale simulated signal (injection) campaign was performed by the LVK detection pipelines over O3 data \citep{GWTC3:2021,sensitivityestimate:2021}. 

For our analysis considering the binary source component masses and redshift, the parameters determining the detectability of signals are the redshifted masses $M_{1,z}$, $M_{2,z}$ and the luminosity distance $D_L(z)$.  Selection effects are thus described via a function $\pdet(M_{1,z}, M_{2,z}, D_L)$ giving the probability of detection for a merger occurring during an observing period; the probability is understood to be marginalised over binary component spins, over extrinsic (location and orientation) parameters, over time of arrival and over detector noise realisations.  

Thus, we require a sufficiently accurate functional approximation for $\pdet$ over masses and distance. Previously, the dependence on masses and spins was modelled via `semi-analytic' functional forms, with corrections derived by various methods in order to fit to injection campaign results \citep{wysocki:2019,talbot:2022}; here we additionally outline an accurate fitting of the distance- or redshift-dependence of $\pdet$, a crucial element in estimating the evolution of BBH merger rates, deferring the demonstration and validation of this method to a dedicated publication \citep{lorenzo:2023}.

We use a physically motivated sigmoid function with a number of free model parameters, fitted by numerically maximising the likelihood of the injection campaign search results.
We define the detection probability $\pdet(M_{1,z}, M_{2,z}, D_L)$ as
\begin{multline}
\pdet(M_{1,z}, M_{2,z}, D_L) = \\ = \frac{\varepsilon_{\max}(M_z)}
  {1 + {\hat{D}_L}^\alpha \exp \left(\gamma(\hat{D}_L - 1) + \delta(\hat{D}_L^2 - 1) \right)}\,,
\end{multline} 
where $M_z = M_{1,z} + M_{2,z}$, and $\hat{D}_L \equiv D_L / D_\mathrm{mid}$ is the luminosity distance scaled by the mass-dependent `midpoint distance' $D_\mathrm{mid}(M_{1,z}, M_{2,z})$, at which the sigmoid falls to 50\% of its maximum value. This sigmoid implements a distance dependence which is universal over all masses, apart from scaling by $D_\mathrm{mid}$.  We fix $\alpha = 2.05$ to fit the theoretical sigmoid data of \citet{finn:1993}\footnote{Allowing $\alpha$ to vary does not improve the overall fit to injection results.}.

The quantity $\varepsilon_{\max}(M_z)$,
\begin{equation}
    \varepsilon_{\max}(M_z) = 1 - \exp(b_0 + b_1M_z + b_2M_z^2)\,,
\end{equation}
represents possible mass dependence of effects that cause the maximum detection probability, even for very loud signals, to fall below unity. In addition, the mass-dependence of the midpoint distance is given by 
\begin{multline}
 D_\mathrm{mid}(M_{1,z}, M_{2,z}) = D_0 \mathcal{M}_z^{5/6} \Big(1 + a_{10}M_z + a_{20}M_z^2 +\\
  + (a_{01} + a_{11}M_z + a_{21}M_z^2) (1 - 4\eta) \Big)\,, 
\end{multline}
where $\eta = (M_{1,z}M_{2,z})/M_z^2$ and the redshifted chirp mass $\mathcal{M}_z = \eta^{3/5}M_z$. 

The sigmoid parameters $\gamma$, $\delta$, $b_0, b_1, b_2$ and the $D_\mathrm{mid}$ parameters $D_0, a_{10}, a_{20}, a_{01}, a_{11}, a_{21}$ are all fitted by maximising the likelihood of the LVK injection results. The motivation for this functional form, and the methodology used for fitting, will be discussed in detail in \citet{lorenzo:2023}. 

We verify that this model is sufficiently complex to accurately describe the BBH detection probability by comparing its predictions for the distribution of detected injections over various physical parameters to the actual injection search results.  We evaluate Kolmogorov-Smirnov (K-S) tests over $\eta$, $M = M_1+M_2$, $M_z$, $\mathcal{M} = \eta^{3/5}M$, $\mathcal{M}_z$, and $D_L$, obtaining K-S statistic values of $0.004$ or less, i.e. a maximum deviation $\leq 0.4\%$ of the actual from the predicted distribution, and $p$-values of $0.09$ or larger, i.e. no significant discrepancies. 

A limitation of our procedure is that we are unable to constrain or verify the behaviour of $\pdet$ outside the physical range of masses and distances covered by the injection set: in extreme cases -- for instance binaries with mass ratio $q$ as small as $\mathcal{O}(0.01)$ -- our model may yield an unphysical estimate. However, its behaviour may readily be checked within the domain where we are estimating BBH merger rates. 

\section{Observed mass ratio distribution}\label{app:qdist}
\begin{figure}
    \centering
    \resizebox{\hsize}{!}{\includegraphics{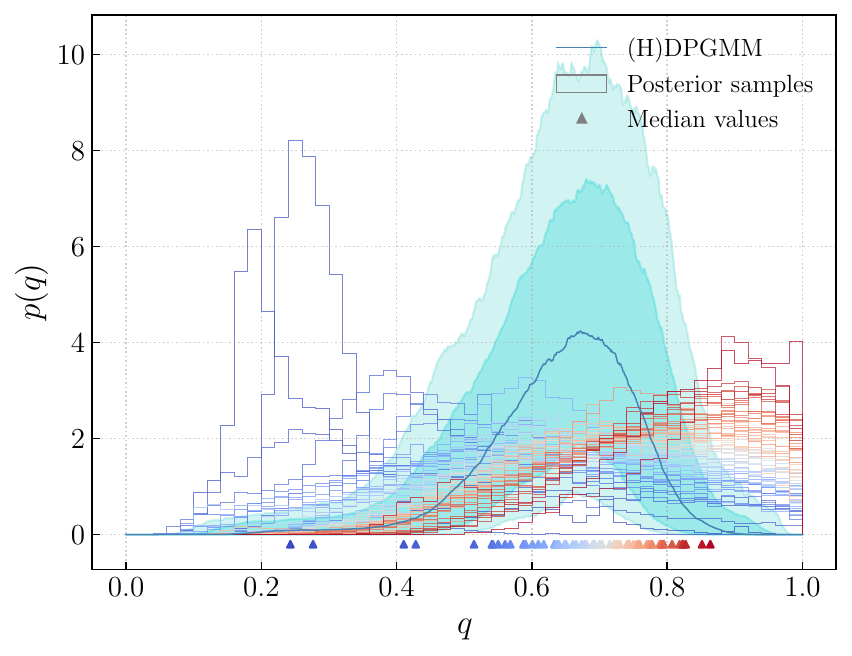}}
    \caption{Marginalised \textsc{figaro} reconstruction of the observed mass ratio distribution (blue solid line -- median -- and shaded regions -- 68\% and 90\% credible regions) along with individual event posterior samples (histograms) coloured according to their median values (triangular markers).}
    \label{fig:qobserved}
\end{figure}

One of the main findings of this paper is that symmetric binaries are largely disfavoured with respect to more asymmetric systems, albeit the majority of individual events being consistent with $q = 1$. In this Appendix, we show how the reconstructed observed probability distribution is compatible with the detected events, deferring a demonstration with simulated data to Appendix~\ref{app:validation}. The observed distribution inferred using GWTC-3 data is reported in Fig.~\ref{fig:qobserved}: this probability density is the marginalised version of the three-dimensional observed distribution $\pobs(M_1,q,z)$ inferred to produce the result presented in the main text of this paper, before correcting for selection effects. The same plot also reports the posterior samples for the individual events included in our analysis as histograms, along with their medians.

We can define consistency with $q = 1$ as $p(q > 0.95)$, finding that 41 events out of 62 have a probability greater than 5\% of having $q > 0.95$. At the same time, however, only 16 events out of 62 have a probability larger than 10\%, and none of them has a probability larger than 20\%\footnote{A more stringent constraint such as $p(q>0.99)$ results in only 4 events being consistent with $q = 1$ with a probability greater than 2\% and none with a probability greater than 5\%.}. In fact, the larger median value among all the posterior distributions is $q = 0.86$, from GW150914. With these things in mind, we see that the inferred observed distribution in Fig.~\ref{fig:qobserved} is consistent with the observations. In particular, the median distribution is consistent with the median values for individual posteriors, with the credible regions accounting for the possible combined fluctuations of the individual posteriors, $q \sim 0.98$. These heuristic arguments suggest that our reconstruction is in agreement with the available observations.

The lack of support for symmetric binaries is then further enhanced by the inclusion of selection effects, which favour asymmetric binaries, and the conversion of the astrophysical distribution into the differential merger rate, leading to the distribution we report in the main text of this paper.

\section{Validation with simulated data}\label{app:validation}
In this section we show that (H)DPGMM and \textsc{figaro} successfully reconstruct a probability density modelled after the findings of \citet{astrodistGWTC3:2023} in $M_1$ and $q$ with as few events as available at the moment.
This is to validate the results presented in this paper and to strengthen the claim that the features we find are due to the actual astrophysical population and not to some artefacts of the model, particularly for what concerns the shape of the mass ratio distribution.

We generated the mock catalogue of GW events using a rejection sampling approach as follows:
\begin{enumerate}[i)]
    \item We drew a set of mock binary parameters from the simulated astrophysical distribution for $M_1$, $q$, and $z$. In particular, the primary mass follows a tapered power-law distribution with a Gaussian peak,
    \begin{multline}
        p(M) = w\frac{(\alpha-1)M^{\alpha}}{M_{\mathrm{min}}^{1-\alpha}-M_{\mathrm{max}}^{1-\alpha}}\qty(1+\mathrm{erf}\qty(\frac{M_{\mathrm{max}}-M}{\lambda_{\mathrm{max}}}))\times \\ \times\qty(1+\mathrm{erf}\qty(\frac{M-M_{\mathrm{min}}}{\lambda_{\mathrm{min}}})) + (1-w) \mathcal{N}\qty(M|\mu,\sigma)\,,
    \end{multline}
    and the mass ratio is distributed according to a power law,
    \begin{equation}
        p(q) \propto q^{\beta}\,.
    \end{equation}
    The redshift distribution is taken to be Gaussian, centred around $\mu_z$,
    \begin{equation}
        p(z) = \mathcal{N}\qty(z|\mu_z,\sigma_z)\,.
    \end{equation}
    This choice for $p(z)$ is dictated by the need to have a model with no significant support in the low-mass, high-redshift region of the parameter space, which is censored by the selection function. A significant support in this region would lead to a bias, as explained in Appendix~\ref{app:censored}.
    We specify the parameters of these distributions in Table~\ref{tab:mock_params};
    \item For each simulated BBH, we evaluated the detection probability $p_\mathrm{det}(M_1,q,z)$ using the selection function by \citet{lorenzo:2023}. For each sample, then, we uniformly drew a number $h$ between 0 and 1: we labelled `detected' all the binaries with $h < p_\mathrm{det}$;
    \item For every detected binary, we generated a set of 2000 posterior samples under the simplifying assumption that the posterior distribution is an uncorrelated multivariate Gaussian distribution in $\log{M_1}$, $\log{\frac{q}{1-q}}$ and $\log{z}$. We made sure that such mock posteriors pass the pp-plot test.
\end{enumerate}

\begin{figure}
    \centering
    \resizebox{\hsize}{!}{\includegraphics{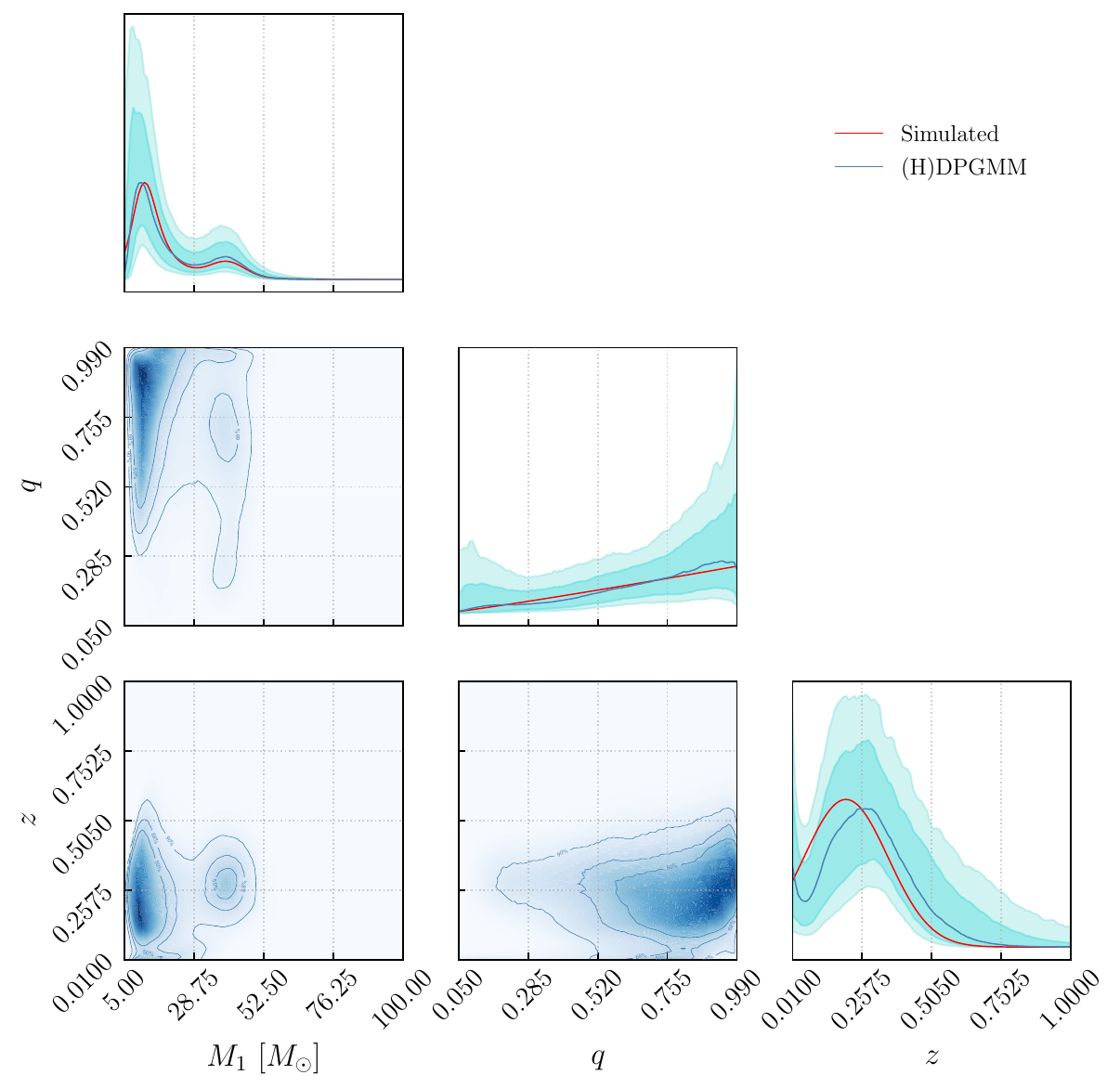}}
    \caption{\textsc{figaro} reconstruction (blue) of the $M_1$, $q$ and $z$ simulated distribution (red). The diagonal panels show the marginal distributions for $M_1$, $q$ and $z$ (top to bottom) respectively and the panels below diagonal show the median distribution marginalised over the third variable.}
    \label{fig:sim_lvk_like}
\end{figure}

\begin{table}
    \centering
    \caption{Parameters of the simulated astrophysical distribution presented in App.~\ref{app:validation}.}
    \begin{tabular}{ll}
    Parameter & Value\\
    \midrule 
    $M_\mathrm{min}$ & $12\ \msun$\\
    $M_\mathrm{max}$ & $70\ \msun$\\
    $\alpha$ & -3.5 \\
    $\lambda_\mathrm{min}$ & $4\ \msun$\\
    $\lambda_\mathrm{max}$ & $3\ \msun$\\
    $w$ & 0.8 \\
    $\mu$ & $40\ \msun$\\
    $\sigma$ & $6\ \msun$\\
    \midrule[0.01em] 
    $\beta$ & 1.1 \\
    \midrule[0.01em] 
    $\mu_z$ & 0.2\\
    $\sigma_z$ & 0.15\\
    \bottomrule
    \end{tabular}
    \label{tab:mock_params}
\end{table}
Our mock catalogue contains 64 events, a number comparable with GWTC-3. Figure~\ref{fig:sim_lvk_like} shows the posterior distribution recovered with \textsc{figaro} after correcting for selection effects, along with the simulated distribution. 
The non-parametric reconstruction is able to capture all the features that are present in the simulated distribution within the 90\% credible region. In particular, the power-law behaviour of the mass ratio distribution is correctly recovered even with as few as 64 event, thus supporting our finding that the astrophysical mass ratio distribution shows little to no support at $q = 1$.

\section{Censored data}\label{app:censored}
Selection effects act as a filter for the astrophysical distribution, modifying the probability of observing a specific binary merger. These effects are due to several reasons: among the others, and without the claim of being exhaustive, we mention the detector noise, the efficiency of the search pipelines and the validity range of waveform models. 

In the framework we applied in this paper, we can correct for selection effects following Eq.~\eqref{eq:selfunc}, under the assumption that we properly sampled all the regions of the parameter space. This assumption does not hold, however, in the case in which the astrophysical distribution has support in a region where the detection probability is negligible\footnote{For example, consider the case in which we expect one detection in a hundred years and we observed for one year only.} or exactly zero, like in a region of the parameter space which is not covered by any waveform model. The selection function would in this case censor a portion of the data.

An astrophysical distribution that extends its support in a censored region would not therefore be reconstructed properly by a non-parametric method: the observed distribution would be cut off by the selection function, biasing the subsequent inference of the astrophysical distribution.

\begin{figure}
    \centering
    \resizebox{\hsize}{!}{\includegraphics{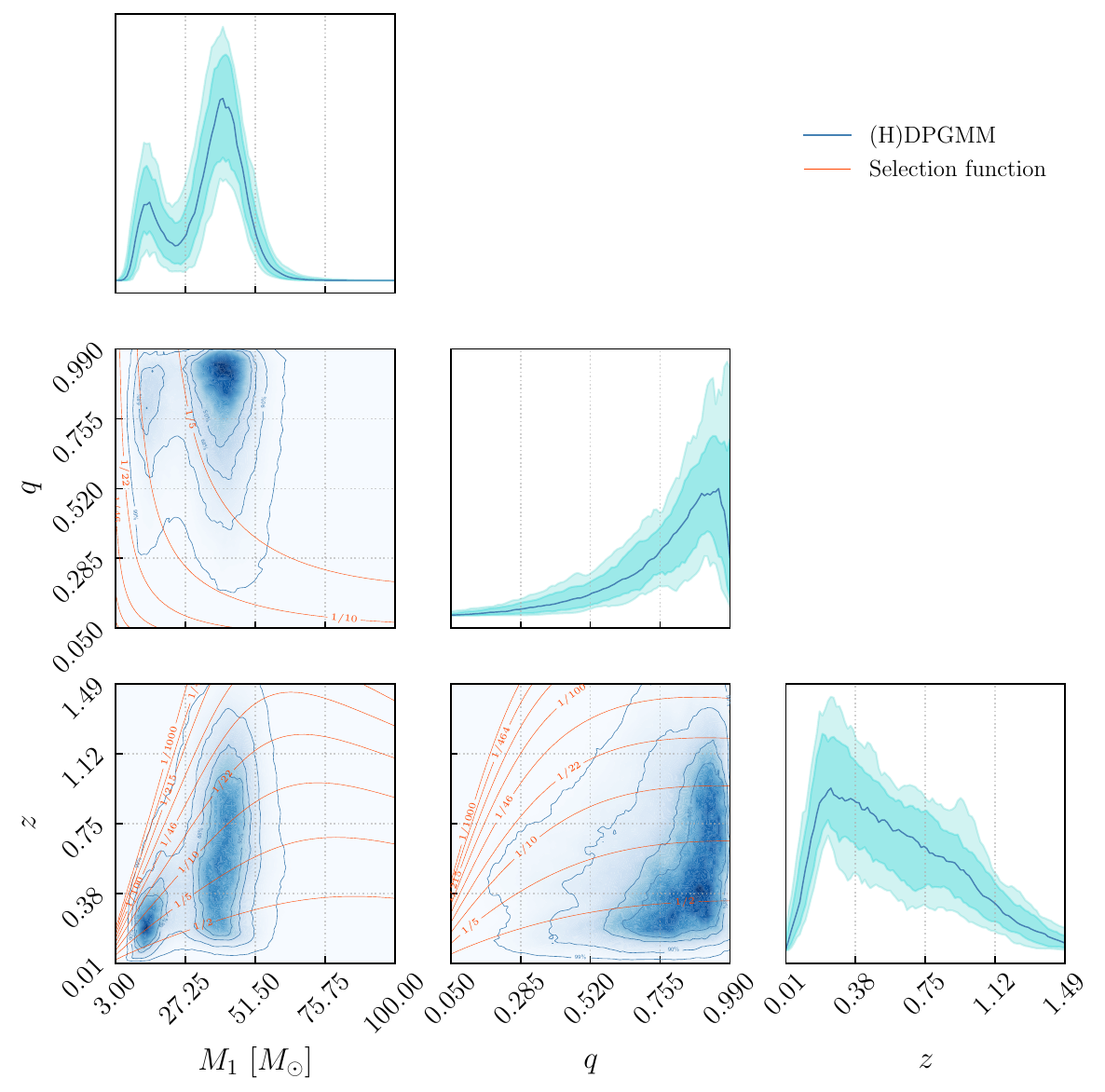}}
    \caption{Observed distribution for the simulated catalogue reconstructed with \textsc{figaro} (blue line) along with the iso-probability levels of the selection function (red). The diagonal panels show the marginal distributions for $M_1$, $q$ and $z$ (top to bottom) respectively and the panels below diagonal show the median distribution marginalised over the third variable.}\label{fig:obs_censored}
\end{figure}

\begin{figure}
    \centering
    \resizebox{\hsize}{!}{\includegraphics{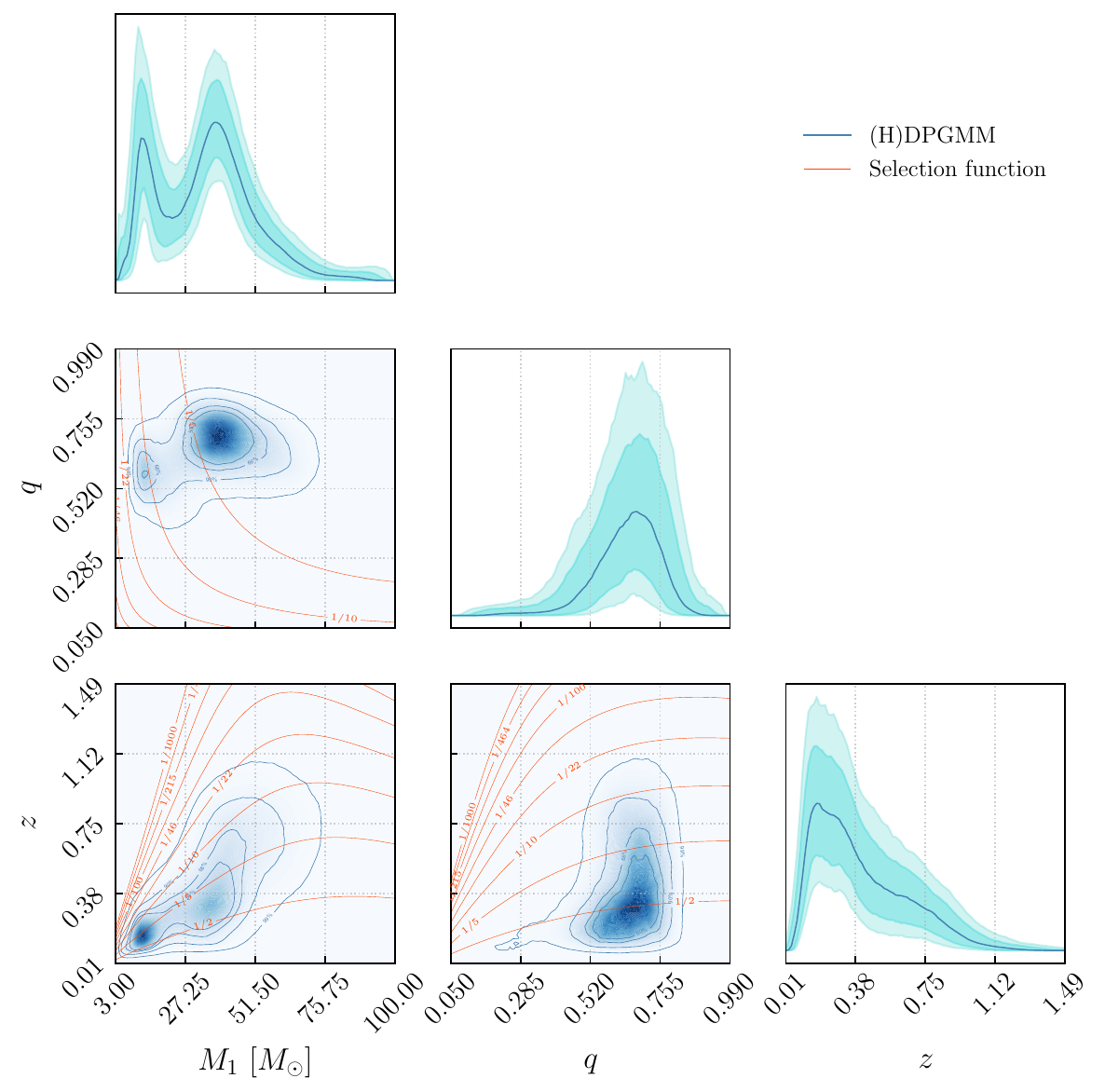}}
    \caption{Observed distribution for GWTC-3 reconstructed with \textsc{figaro} (blue line) along with the iso-probability levels of the selection function (orange). The diagonal panels show the marginal distributions for $M_1$, $q$ and $z$ (top to bottom) respectively and the panels below diagonal show the median distribution marginalised over the third variable.}\label{fig:observed_distribution}
\end{figure}

In Fig.~\ref{fig:obs_censored} we show the behaviour of an observed distribution whose astrophysical distribution has support in a censored region of the selection function used throughout this paper \citep{lorenzo:2023}. The simulated underlying distribution is the same as in Appendix~\ref{app:validation}, with the exception of $p(z)$:
\begin{equation}
    p(z) \propto \frac{\dd V_\mathrm{c}}{\dd z}\,.
\end{equation}
The median two-dimensional $M_1-z$ distribution follows, in the low-mass, high-redshift region, the iso-probability curves of the selection function, thus suggesting the censoring of a portion of the data. 
Conversely, this behaviour is not present for the credible regions of the observed distribution reconstructed with GWTC-3 (reported in Fig.~\ref{fig:observed_distribution})  suggesting, although heuristically, that we are not in presence of censored data.
\end{appendix}
\end{document}